\documentclass{article}

\usepackage[12pt]{extsizes}
\usepackage[super,sort&compress,comma]{natbib} 
\usepackage[version=3]{mhchem}
\usepackage{mathptmx}
\usepackage{sectsty}
\usepackage{graphicx} 
\usepackage{lastpage}
\usepackage[format=plain,justification=justified,singlelinecheck=false,font={stretch=1.125,small,sf},labelfont=bf,labelsep=space]{caption}
\usepackage{float}
\usepackage{fancyhdr}
\usepackage{fnpos}
\usepackage[english]{babel}
\addto{\captionsenglish}{%
  
}
\usepackage{array}
\usepackage{droidsans}
\usepackage{charter}
\usepackage[T1]{fontenc}
\usepackage[usenames,dvipsnames]{xcolor}
\usepackage{setspace}
\usepackage[compact]{titlesec}
\usepackage{hyperref}

\usepackage{comment}
\usepackage{amsmath}
\usepackage{amsfonts}
\usepackage{ragged2e}

\usepackage{physics}

\usepackage[a4paper, headheight=0.5cm, margin=2.5cm]{geometry}

\begin{document}

\centering\noindent \Large{\textbf{{Investigating Discontinuous X-ray Irradiation as a Damage Mitigation Strategy for \ce{[M(COD)Cl]2} Catalysts}}}

\vspace{7.5mm}

\noindent\large{Nathalie K. Fernando,$^{\ast}$\textit{$^{a}$} Claire A. Murray,\textit{$^{b}$} Amber L. Thompson,\textit{$^{c}$} Katherine Milton,\textit{$^{a}$} Andrew B. Cairns,\textit{$^{d}$} and Anna Regoutz\textit{$^{a,e}$}}  \\

\vspace{5mm}
\noindent\small\textit{{$^{a}$~Department of Chemistry, University College London, 20 Gordon Street, London, WC1H~0AJ, UK.}\\
\textit{$^{b}$~Diamond Light Source Ltd, Diamond House, Harwell Science and Innovation Campus, Didcot, Oxfordshire, OX11~0DE, UK.}\\
\textit{$^{c}$~Chemical Crystallography, Chemistry Research Laboratory, University of Oxford, South Parks Road, Oxford OX1~3QR, UK.}\\
\textit{$^{d}$~Department of Materials, Imperial College London, Royal School of Mines, Exhibition Road, SW7~2AZ, UK.}\\
\textit{$^{e}$~Present address: Department of Chemistry, Inorganic Chemistry Laboratory, South Parks Road,OX1~3QR, UK.}}\\

\vspace{10mm}
\justifying
With the advent of ever more intense and focused X-ray sources, including in laboratories, at synchrotrons, and at X-ray free electron lasers, radiation-induced sample change and damage are becoming increasingly challenging. Therefore, the exploration of possible mitigation strategies is crucial to continue to allow the collection of robust and repeatable data. One mitigation approach is the introduction of short, X-ray-free ``dark'' periods. However, it is unclear whether this strategy minimises damage or, in actuality, promotes it through a phenomenon called ``dark progression'', i.e. the increase or progression of radiation damage that occurs after the X-ray beam is turned off. This work discusses the influence of introducing dark periods and their duration on the radiation-induced changes in two model small-molecule catalysts, \ce{[Ir(COD)Cl]2} and \ce{[Rh(COD)Cl]2}, exposed to X-ray radiation in powder diffraction (PXRD) and photoelectron spectroscopy (XPS) experiments. This provides, for the first time, insights into how damage progresses under varying radiation regimes and allows the distinction between the processes that affect the unit cell itself, the individual molecular units, and the respective atomic chemical environments. Furthermore, it provides the basis for informed decision-making in the design of future experiments where the need to minimise radiation-induced damage is crucial. \\

%\footnotetext{\dag~Electronic Supplementary Information (ESI) available: [details of any supplementary information available should be included here]. See DOI: 10.1039/cXCP00000x/}

%%%MAIN TEXT%%%%
\justifying
\section{Introduction}

In recent years, X-ray free electron lasers have emerged as a modern, fourth-generation X-ray source capable of probing samples with ever-increasing resolution, not previously achievable. Such short, high-impulse systems have been found to outrun timescales of sample damage and have paved the way for the modern experimental approach of acquiring high-resolution data before sample quality is compromised. In diffraction experiments, this is commonly referred to as ``diffraction before destruction'' and is the basis of typical serial crystallography experiments, wherein a single flash of X-ray exposure gives rise to a diffraction pattern before the sample is replenished to obtain more undamaged patterns. Although such techniques are now a routine method of sample damage mitigation, access to such large-scale facilities is not always feasible or possible. As such, only in the past decade have a handful of studies emerged aiming to apply the theory of outrunning damage in protein crystals used in XFELs to more accessible, conventional XRD facilities. This work explores, for the first time, the impact of introducing short X-ray-free periods into small molecule X-ray characterisation of \ce{[Ir(COD)Cl]2} and \ce{[Rh(COD)Cl]2} to answer whether this hinders sample damage or promotes it. \par

The increase or progression of radiation damage that occurs after the X-ray beam is turned off is often termed ``dark progression''. The handful of studies that exist on this topic, and which will be discussed below, investigated dark progression to probe damage timescales and determine a potential dose-rate dependence of global radiation damage, as well as the possibility of outrunning it. Despite these reports, there remains a noticeable gap in understanding dark progression in small molecule systems, which are typically subject to direct X-ray exposure without a parent solvent, in contrast to protein crystallography. The aim of this study is to thus bridge this knowledge gap and determine the influence of dark periods on radiation damage effects in these metal complexes.\par

It is first important to revisit briefly the timescales of radiation-induced processes. It is well-understood that primary X-ray photon-electron interactions occur on timescales of femtoseconds. These interactions often result in the formation of photoelectrons, which in turn interact with atoms comprising the sample and produce free radicals, which are known to occur on sub-picosecond timescales. Photoelectron-electron (matter) interactions also occur on femtosecond timescales. These excitations initiated by the X-ray photon thermally diffuse, resulting in global sample damage. The rate at which thermal diffusion occurs is very much temperature-dependent, slowing at lower temperatures. Most small-molecule X-ray experiments are conducted at room temperature, where the diffusion rate is high, potentially leading to bond damage. This can lead to further excitation reactions of even longer duration.\par 

Anecdotal evidence of the degradation of protein crystals at room temperature after being removed from an X-ray beam was first reported by Blundell and Johnson in 1976.~\cite{Blundell1976} Subsequent investigations into this phenomenon have not always been able to corroborate these findings, fueling greater uncertainty. At cryotemperatures (T=100~K), Ravelli and McSweeney found no conclusive evidence of dark progression with respect to the global metrics, mosaicity and B-factors, an indicator of crystal disorder and atomic fluctuation, respectively.~\cite{Ravelli2000} However, the authors noted an increase in the site-specific R\textsubscript{merge} value, used to describe the agreement between multiple measurements of a given reflection (to determine data quality), during X-ray-free periods of 10~min and 19~hours. Work by Kmetko and Southworth-Davies also showed no evidence of dark progression at T=100~K and T=300~K, using global damage metrics.~\cite{Kmetko2006, Davies2007} Beyond typical global damage metrics, using a UV/Vis microspectrophotometer device inline during XRD experiments, McGeehan~\textit{et al.}\ noted a drop in optical density of proteins during X-ray-free dark periods, signalling the site-specific decay of disulfide radicals and trapped electron content at temperatures of 100, 130 and 160~K.~\cite{Mcgeehan2009} \par 
 
By the early 2000s, no concrete evidence of dark progression of global damage had been established at either T=100~K or T=300~K. The focus then shifted to exploring dose rate as a possible contributing factor to dark progression. For instance, Warkentin \textit{et al.}\ studied dark progression in thaumatin crystals by conducting impulse-response measurements at different dose rates.~\cite{Warkentin2011} The increase in B-factors, measured straight after dark intervals (of 240 and 660~s), determined the extent of dark progression. Contrary to earlier reports, dark progression was observed on timescales from 200 to 1200~s between temperatures of T=180~K and T=240~K. For instance, a 27\% reduction in global damage at T=240~K was reported in datasets collected in 600~s, compared to those collected during 1200~s long measurements, suggesting a reduction in damage with increasing dose rate. However, no dark progression was observed from T=25~K to T=180~K. The authors also found that longer dark intervals result in greater global damage. This study showed the potential for outrunning some global radiation damage at temperatures above T=180~K by combining several factors, such as faster data collection and moderate cooling.\par

The authors noted that the shortest damage timescale that can be determined experimentally is limited by the duration of the probe pulse, which is defined by the time it takes to obtain a diffraction pattern of high enough resolution to assess the damage. If the timescales for damage processes are significantly longer than the experimental time or much shorter than the probe pulse, no dark progression will be measured. In a follow-up study, they tested a wider range of dose rates. They noted that the dose-rate effect of radiation damage is too small to be noticeable at T=300~K.~\cite{Warkentin2012} Thus suggesting that at T=300~K, radiation damage processes occur faster than the approximately 1--10~s minimum data collection time. Warkentin~\textit{et al.}\ also hypothesised that adding X-ray-free pauses during measurement contributed to the inverse dose-rate effect observed via relaxation processes.  \par 

In the same year, Owen~\textit{et al.}\ studied the room temperature X-ray-induced decay of protein and virus crystals as a function of dose rate with faster readout detectors and short X-ray exposure times using the full flux of an undulator beamline at Diamond Light Source.~\cite{Owen2012} Two detector-readout modes were compared to determine whether continuous diffraction data collection (without repeated pauses in X-ray irradiation while the detector reads out) results in increased crystal lifetimes. The authors showed that crystal lifetime was reduced by a factor of two when there was a pause between frames within a dataset. This is in contrast to the findings in the 2012 study by Warkentin~\textit{et al.}~\cite{Warkentin2012}

It has been reported previously that very short femtosecond XFEL pulses deposit a massive X-ray dose, leading to indexable diffraction from crystals at RT before the complete ionisation of the structure destroys it in less than 100~fs.~\cite{Neutze2000, Chapman2011} Based on these observations, Owen~\textit{et al.}\ suggested three sources for radiation damage in macromolecules at RT and their significance at slow and fast timescales. On slow timescales ($>$1~s), radical diffusion (and quenching within the solvent) occur, while on fast timescales ($<$60~ms), radical diffusion, quenching, and recombination are all relevant. Based on these findings, Owen~\textit{et al.}\ suggest that more intense beams and faster detectors might make RT data collection more appealing for MX studies.

The progression of damage during dark periods can be attributed to the timescales of diffusion and the reaction of free radicals, as well as diffusive conformational motions of the structure. Dark progression is also attributed to the mobility of molecular hydrogen gas, which is generated by the X-ray interaction with the sample and subsequently trapped within the crystal. Meents~\textit{et al.}\ attribute the temperature dependence of radiation damage at low temperatures in protein crystals to the temperature dependence of hydrogen gas diffusion.~\cite{Meents2010} Hydrogen gas produced inside the crystal can perturb the steady state by creating pressure, which can be released or redistributed within the crystal. This may be a source of the radiation-induced bond-breaking and creation of defects in crystals. Local atomic displacements and electronic rearrangements create further pressures, leading to lattice expansion. Protein crystals typically comprise mostly of hydrogen atoms, and therefore, it is expected that a significant fraction of broken bonds will result in the release of hydrogen. Therefore, hydrogen contributes to the internal pressure of the crystal and consequently global radiation damage, particularly at room temperature, where molecular hydrogen is highly mobile. As expected, hydrogen mobility does not contribute to diffraction damage effects at cryotemperatures.\par

Unlike the previous studies of biological macromolecular crystals outlined above, here, the possible occurrence of dark progression at RT, as noted by Owen~\textit{et al.}, is investigated in model small-molecule systems, namely \ce{[Ir(COD)Cl]2} and \ce{[Rh(COD)Cl]2}, where COD = 1,5-cyclooctadiene. This is achieved by probing global (XRD) and specific (XPS) changes. Whether the duration of the dark period in PXRD experiments influences the extent of global damage is also explored.

\section{Methods}
To determine the influence of increasing X-ray-free, dark periods during long-duration X-ray experiments, the model complexes, \ce{[Ir(COD)Cl]2} and \ce{[Rh(COD)Cl]2} (Sigma Aldrich, reference IDs 683094 and 227951, and purity levels of 97\% and 98\%, respectively) were investigated. Firstly, synchrotron-based PXRD experiments were conducted to compare the global effects on the molecular unit and overall crystal structure of introducing dark periods. Next, laboratory-based X-ray photoelectron spectroscopy experiments were carried out in a method comparable to an initial radiation damage study by the authors of this work,~\cite{Fernando2021}, to probe potential local chemical changes to the system, invisible to diffraction. 

\begin{figure}[h!]
\centering
\includegraphics[width=0.65\textwidth]{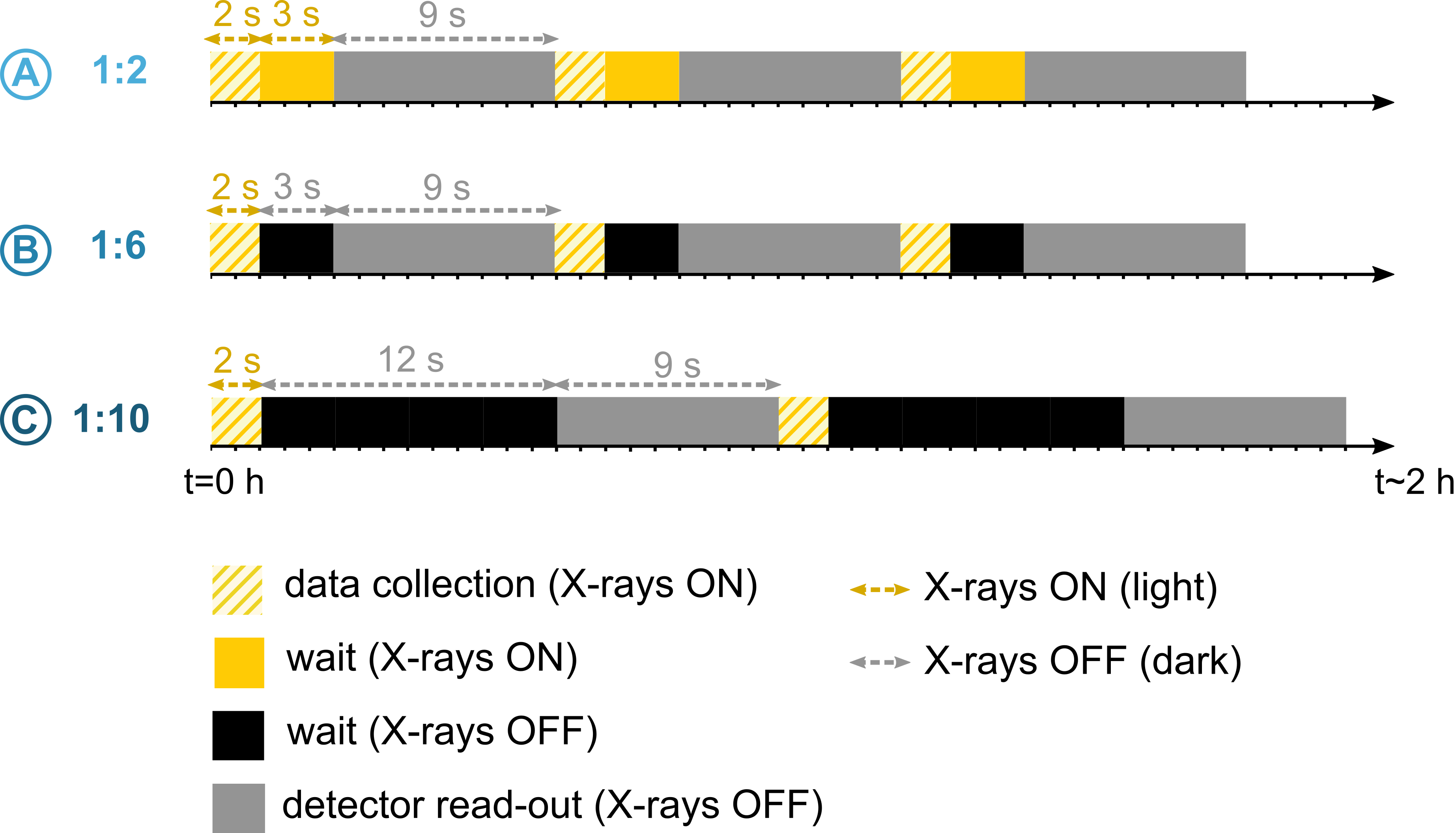}
\caption{\label{fig:gap_method} Schematic outlining the experimental PXRD procedure of the three X-ray irradiation regimes first comparing the X-ray on to X-ray off durations of 1:2 (A) and 1:6 (B), to study the influence of a 3~s X-ray free window compared with an irradiated equivalent time. Then, comparing the 1:6 (B) and 1:10 (C) regimes, where the X-ray off wait window has been extended from 3 to 12~s. The scans were repeated until the total elapsed experimental time was approximately 2~h per irradiation regime.}
\end{figure}

For the PXRD experiments at beamline I11, Diamond Light Source, UK \ce{[Ir(COD)Cl]2} and \ce{[Rh(COD)Cl]2} were mounted into 0.3~mm borosilicate capillaries. The PXRD study can be separated into two components, the first being a comparison of incorporating a light (X-ray beam on) period and a dark (X-ray beam off) period of the same duration, represented in Figure~\ref{fig:gap_method}.

In the first instance, the standard experimental protocols used at beamline I11 were followed, namely a 2~s data collection time, followed by 3~s of irradiation, and finally 9~s of detector read-out/data saving time, during which time the X-ray shutter was in, and the sample was no longer irradiated. This irradiation regime corresponds to an irradiation:dark time ratio of 1:2 per pulse. This process was repeated in iterations such that 500 diffraction patterns were collected in total (over 2~hours). As in ~\cite{Fernando2022}, 0.3~mm borosilicate capillaries were used with a photon energy of 15~keV throughout at T=300~K. A second data set was collected in the same way, except for the 3~s irradiation time being replaced with 3~s of X-ray-free, dark time to determine the impact of incorporating an X-ray-free element into data collection. This regime corresponds to an irradiation:dark time ratio of 1:6 per pulse.\par

The second part of the PXRD study aimed to determine the effect of the duration of this dark period on global structure. To compare gap durations, the dark period between the 2~s data collection and 9~s detector read-out/data saving time was increased from 3~s to 12~s dark period (1:10 irradiated:dark ratio). Again, these diffraction patterns were collected iteratively so that 500 total scans were obtained at the end of the experiment. A combination of batch Le Bail and Rietveld PXRD refinements was carried out on this data using the TOPAS Academic v7 software package.~\cite{Coelho2018, TOPAS7} 
In order to determine changes to the individual lattice parameters and overall unit cell volume over the measurement period.~\cite{Dinnebier2018} For batch Rietveld refinements,~\cite{Rietveld1967} starting atomic coordinates were taken from results of the single crystal XRD structure solution conducted by Fernando~\textit{et al}.{\cite{Fernando2021} M--Cl and C--C distance restraints were applied to achieve a reasonable starting structure and were maintained for the rest of the dataset.\par

For ease of comparison across the structures corresponding to each irradiation regime during structural refinements, a rigid-body approach to describe the cyclooctadiene (COD) ligands was used, as first introduced in ~\cite{Fernando2022}.\par 

To facilitate the comparison of damage of both the Ir and Rh catalyst at the different light/dark ratios, X-ray dose was estimated using the RADDOSE-3D tool.~\cite{Zeldin2013, Bury2018} As with the dose estimations carried out in Fernando~\textit{et al.} 2021 and Fernando~\textit{et al.} 2022,~\cite{Fernando2021, Fernando2022} the small molecule crystal addition to the programme, introduced by Christensen~\textit{et al.}, was used, along with the capillary model introduced by Brooks-Bartlett \textit{et al.}~\cite{Christensen2019, Brooks2017} The beam size was obtained from beam characterisation of the I11 beamline undertaken by Thompson~\textit{et al.}~in 2009 at 15~keV, see Figure~2(a) and (b) in Ref.~\citenum{Thompson2009}.\par

To complement PXRD findings on global changes, XPS experiments on the Thermo Scientific K-Alpha\textsuperscript{+} instrument at Imperial College London, UK, were conducted to determine the effect of introducing dark periods on specific changes. The spectrometer contains a microfocused monochromated Al K$\alpha$ X-ray source ($h\nu$=1.487~keV), with a $180^\circ$ double focusing hemispherical analyser and 128 channel detector. The spectrometer operates with a 6~mA X-ray anode emission current and 12~kV accelerating voltage and has a spot size of 400~$\mu$m. The energy resolution of the spectrometer was determined to be 0.44~eV. The photon flux at these experimental parameters is calculated to be approximately 3.8$\times10^{10}$~photons/s from source parameters obtained from the manufacturers. The base pressure of the spectrometer was 2$\times10^{-9}$~mbar.\par
As a control dataset, each sample, mounted on adhesive carbon tape, was irradiated continuously for a total of 13~hours, with core level (CL) and valence band (VB) spectra being collected iteratively every 2 hours. For the dark measurement, on a fresh sample, two scans of each CL and VB (totalling approximately 3~min) were collected. The X-ray beam position was then moved away from the irradiated spot on the sample for 1~min (in reality, including data saving and sample height alignment, this was 1~min 50~s.) The above steps comprised a single iteration. 50 iterations were measured to obtain comparable spectral resolution to the continuous dataset. These 50 iterations comprised a single group and were repeated eight times, with the total X-ray irradiation time per group being approximately 2~hours. XPS peak fit analysis was carried out using the ThermoScientific Avantage software package, following the method outlined in Fernando~\textit{et al.}\cite{Fernando2021}. The peak fitting functionality was used, with the in-built \textit{Smart} background, a development of the Shirley background.~\cite{Watts2019} The average published spin-orbit splittings of Ir, Rh, and Cl from the NIST X-ray Photoelectron Spectroscopy database informed the constraints applied on the peak positions. The FWHM parameter and Lorentzian/Gaussian ratios of the Voigt function used to fit the peaks were refined, and the areas of each spin-orbit peak were constrained according to their Scofield photoionisation cross sections.~\cite{Scofield_1973, Dig_Sco_2020} 

Following on from the first application of RADDOSE-3D in XPS studies, presented in Fernando~\textit{et al.}\ 2021, the tool was again used to estimate dose over the long course of X-ray exposure during the XPS experiments outlined here. Dose is estimated using the AD-WC (Absorbed Dose - Whole Crystal) metric, defined as the total energy absorbed divided by the mass of the whole crystal. This metric is chosen since the FWHM of the X-ray beam is such that the entirety of individual crystals is irradiated by the beam. A tabulated summary of the main input parameters for \ce{[Ir(COD)Cl]2}, \ce{[Rh(COD)Cl]2} used in all RADDOSE-3D calculations are provided in Table~S1 in the Supplementary Information.\par

Given each dark:light irradiation regime will have a unique maximum X-ray dose, in order to more effectively compare crystallographic and chemical changes across the regimes, all analysis is carried out on the changes as a function of X-ray dose i.e. normalised to the dose. 

\section{Results \& Discussion}

The results of the experiments outlined can be split into three parts. First, comparisons between unit cell behaviour under the three irradiation regimes in PXRD are explored. Second, atomic distortions of the central M--Cl core are discussed. Finally, results from complementary XPS experiments are explored to determine whether dark progression can be observed for such local changes.

\subsection{Changes to the Unit Cell}

Initial full profile (Le Bail) refinements provide insights into the variability of unit cell change across the three regimes studied (1:2, 1:6 and 1:10 irradiated to dark ratio).~\cite{LeBail1988} First considering \ce{[Ir(COD)Cl]2}, in Figure~\ref{fig:Ir_2_6_10_lattice_params}, it is evident that all three lattice parameter lengths undergo expansion. The order of expansion is not only consistent with observations from a previous study by some of the authors,~\cite{Fernando2021} but also consistent across the three irradiated:dark duration ratios, with the $c$ axis experiencing the smallest increase, then $a$ closely followed by $b$, which shows the greatest increase.\par 

\begin{figure*}[h!]
\centering
\includegraphics[width=\textwidth]{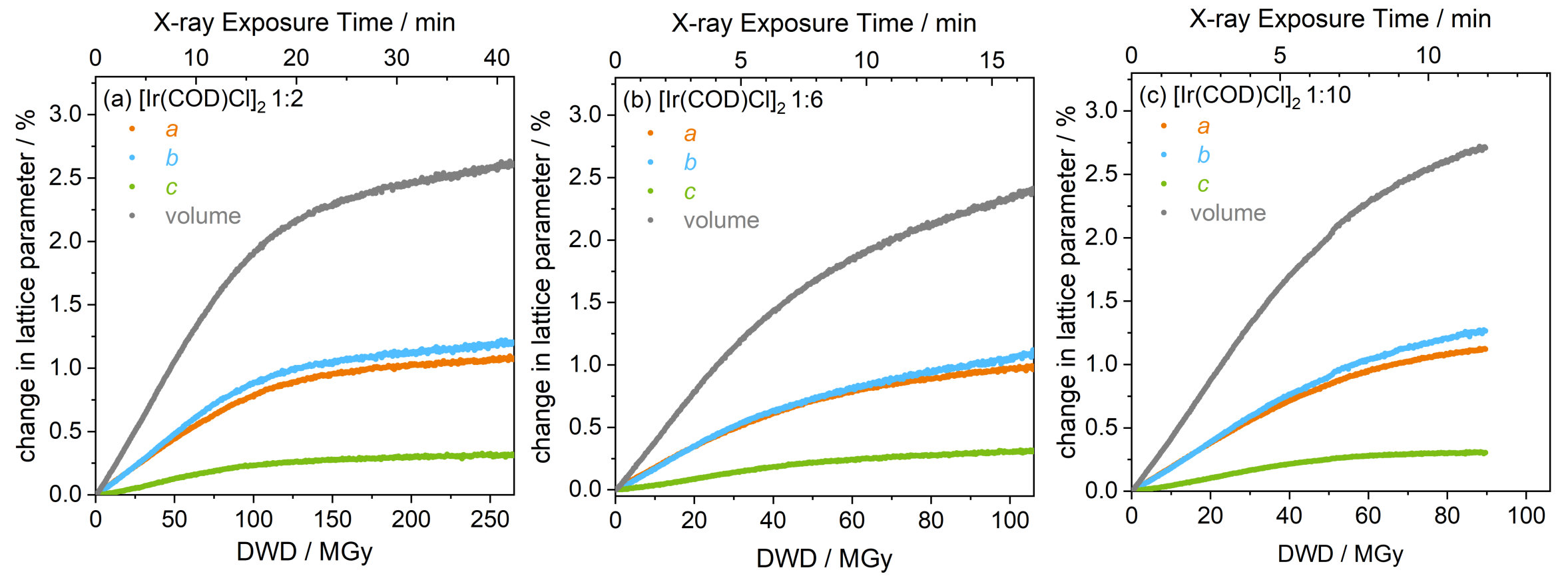}
\caption{\label{fig:Ir_2_6_10_lattice_params}Quantitative results showing the change in lattice parameters and unit cell volume of \ce{[Ir(COD)Cl]2} as a function of diffraction-weighted dose (DWD), extracted from Le Bail refinements of the PXRD data for ratios of irradiation to X-ray free (dark) durations of (a) 1:2, corresponding to 2~s data collection, 3~s of further irradiation and 9~s of data saving (dark time), (b) 1:6, corresponding to 2~s data collection, 3~s dark and a further 9~s of data saving and (c) 1:10, which is 2~s data collection, followed by 12~s of dark and an additional 9~s for data saving. All plots include both the X-ray exposure time as well as the calculated X-ray dose from RADDOSE-3D. From error propagation calculations, the approximate errors in (a) are $\pm$0.007\% and $\pm$0.01\% for the percentage change in lattice parameter and volume, respectively. Similarly for (b), these values are $\pm$0.03\% (lattice parameters) and $\pm$0.06\% (volume). For (c), the approximate errors are $\pm$0.007\% (lattice parameters) and $\pm$0.02\% (volume).}
\end{figure*}

By directly comparing the changes to the individual lattice parameters of the Ir complex, both as a function of dose and the three different irradiation regimes outlined above, notable consistencies across all lattice parameters, $a$, $b$, $c$ and cell volume can be observed (see Figure~S1 in the Supplementary Information for comparative plots of all four). The 1:2 regime shows the smallest increase in lattice parameters, followed by 1:6 and finally 1:10, which shows the greatest increase as a function of dose. For all lattice parameters, the linear expansion between the 1:6 (12~s gap) and 1:10 (21~s gap) datasets appear to be identical from the onset of irradiation until a cumulative dose of approximately 20~MGy after which point the curves diverge. The difference in the onset of curve plateauing is also evident, with the 1:2 dataset plateauing later than the two longer dark-duration datasets. Results from Le Bail refinements suggest that there is a clear dependence of unit cell expansion, on the X-ray irradiation regime chosen. A longer proportion of X-ray irradiation relative to dark time is clearly more favourable in these experimental timescales, in which long-scale diffusion processes dominate. Conversely, in XFELs, for instance, the effects of even very short-scale damage processes are limited.~\cite{Warkentin2012} This suggests that reducing the light-to-dark ratio enables long-range damage processes to dark progress further through the crystal. In other words, reducing the duration of the X-ray-free period means the time between measurements is shorter, and as such, only a fraction of the dark progression of long-range damage is observed in the subsequent diffraction pattern.  

In order to quantify the point at which the unit cell volume of the Ir complex begins to reach saturation, the onset of the plateau observed from the above Le Bail refinements is determined from the first and second derivative curves of volume as a function of X-ray dose, see Figure~S2 in the Supplementary Information. From this, it is then possible to determine whether a certain X-ray irradiation/dark regime promotes the crystal to reach maximum volume earlier with respect to the others for the same cumulative absorbed dose. 

By definition, the region defined by the point at which the linear region diverges, signals the onset of the plateau to the point at which it begins to stabilise in the first derivative curve. The point of inflexion of this change can be seen in the second derivative curve, which is used to define the point of plateau. The first derivative curve for all three irradiation-dark regimes, see Figure~S2(b) in the Supplementary Information, can be separated into three main regions. There is an initially flat (or subtle upwards incline) from the onset of irradiation up to a certain dose. This dose is 30.9~MGy, 10.8~MGy and 16.7~MGy, for the 1:2, 1:6 and 1:10 datasets, respectively. After this critical dose is reached, there is a significant downward slope, signalling the plateau region, stabilising at a later cumulative dose. This dose is 221.3~MGy and 100.0~MGy for 1:2 and 1:6, respectively. The 1:10 first derivative of volume is too subtle to notice the point at which the curve flattens, meaning that the plateau has not yet stabilised in the timeframe of the experiment.\par 

Considering both Figures~S2(b) and (c) in the Supplementary Information, the point of plateau onset and the mid-point are almost the same for both 1:6 and 1:10 datasets, the difference between which could be considered insignificant. However, for the 1:2 dataset, with the largest irradiated-to-dark duration ratio, the plateau region occurs at a much higher dose, meaning that the unit cell volume increase reaches saturation at a similar point regardless of the duration of the dark period. In other words, the difference between the longer-duration dark X-ray regimes and the light 1:2 regime is stark. This potentially suggests that the long-range damage time scales roughly correspond to the 1:6 dark duration (3~s dark gap $+$ 9~s dark data saving time) such that the dark progression reaches its maximum in the 12~s when the X-ray shutter is on. At the 1:2 regime, it is hypothesised that the diffraction pattern collection times are so frequent that it is too early in the long-range damage process with respect to the other two regimes. \par

\begin{figure*}[ht!]
\centering
\includegraphics[width=\textwidth]{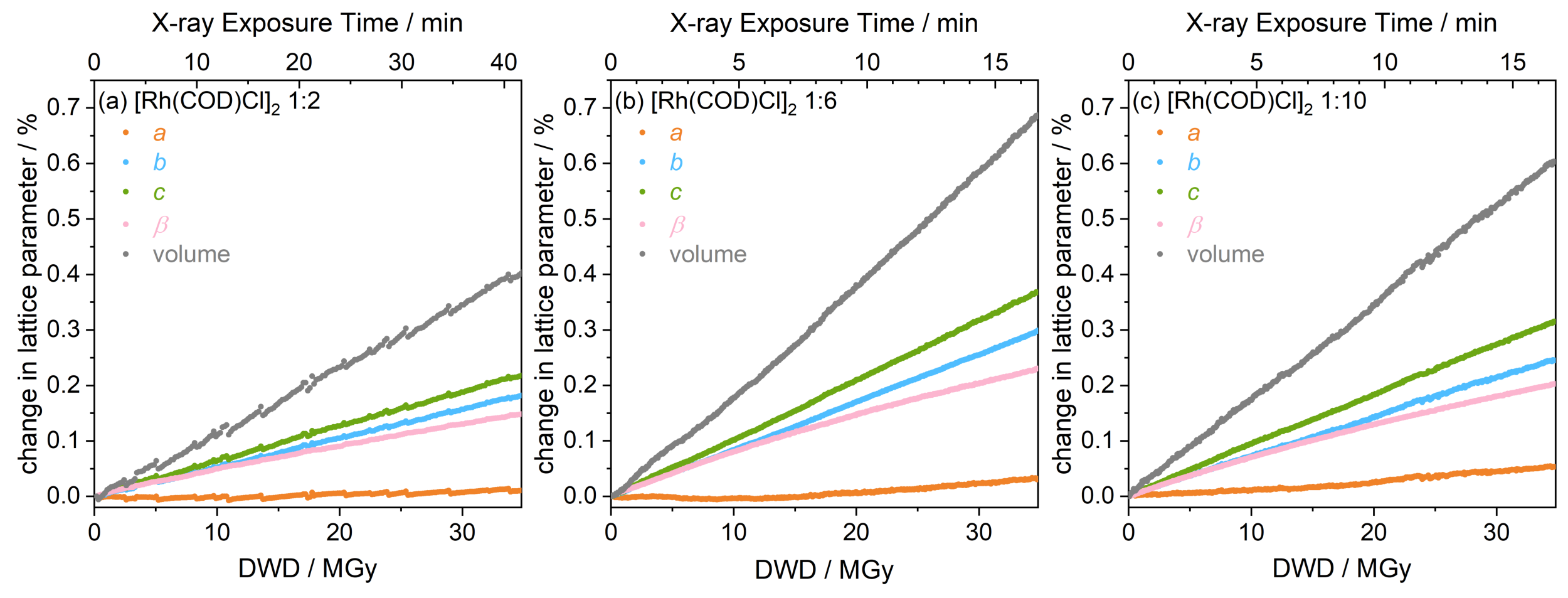}
\caption{\label{fig:Rh_2_6_10_lattice_params}Quantitative results showing the \ce{[Rh(COD)Cl]2} change in lattice parameters and unit cell volume as a function of diffraction-weighted dose (DWD), extracted from Le Bail refinements of the PXRD data for ratios of irradiation to X-ray free (dark) duration of (a) 1:2, corresponding to 2~s data collection, 3~s of further irradiation and 9~s of data saving (dark time), (b) 1:6, corresponding to 2~s data collection, 3~s dark and a further 9~s of data saving and (c) 1:10, which is 2~s data collection, followed by 12~s of dark and an additional 9~s for data saving. All plots include both the X-ray exposure time as well as the calculated X-ray dose from RADDOSE-3D. From error propagation calculations of the last data point, the approximate errors in (a) are $\pm$0.004\%, $\pm$0.001\% and $\pm$0.01\% for the percentage change in lattice parameter, $\beta$, and volume, respectively. Similarly for (b), these values are $\pm$0.001\% (lattice parameters), $\pm$0.001\% ($\beta$) and $\pm$0.003\% (volume). For (c), the approximate errors are $\pm$0.002\% (lattice parameters), $\pm$0.001\% ($\beta$) and $\pm$0.003\% (volume).}
\end{figure*}

The same approach of Le Bail refinements was carried out on the equivalent monoclinic \ce{[Rh(COD)Cl]2} datasets to determine changes to lattice parameters $a$, $b$, $c$, $\beta$ and volume, see Figure~\ref{fig:Rh_2_6_10_lattice_params}. As with the Ir complex, the Rh complex shows the same order of lattice parameter increase across the three different regimes ($a<<b<c$). As reported in Ref.~\cite{Fernando2021}, the overall changes are almost an order of magnitude smaller with respect to the unit cell changes observed in the Ir complex. At 34.8~MGy, the unit cell volume expands by approximately 0.4~\%, 0.7~\% and 0.6~\% for the 1:2, 1:6 and 1:10 setups, respectively. Despite these very subtle changes, comparing first the light vs. dark cases, it is clear to see that in the 1:2 light-to-dark regime, the rate of increase across all refined lattice parameters, is lower than the `darker' 1:6 regime. The changes are particularly small in the 1:2 case as evidenced in Figure~\ref{fig:Rh_2_6_10_lattice_params}(a), by the periodic discontinuities in the linear trend which can be attributed to the electron top-up of the synchrotron storage ring. However, unlike in the Ir complex, the lattice parameter increases do not appear to have reached a plateau at the final cumulative dose corresponding to the end of the experiment--- in other words, a linear increase is observed throughout. Considering the effect of dark gap duration across all lattice parameters, the 1:6 (12~s gap) case increases at a greater rate than the 1:10 (21~s gap) case, although, quantitatively, the margins by which they differ are minimal.\par

As in the Ir complex case, to better compare differences across the three measurement regimes, each lattice parameter was plotted separately as a function of X-ray dose and irradiation-gap regime, see Figure~S3 in the Supplementary Information. Comparing first the 1:2 against the 1:6 regime, it is clear that all refined lattice parameters of the 1:2 increase at a notably lower rate than those of the 1:6, and as such, it appears to be the more favourable of the settings to limit radiation-induced structural damage. This is consistent with the findings of the lattice parameter increases of the \ce{[Ir(COD)Cl]2}. Next, comparing the 1:6 against the 1:10 data to determine the effects of gap duration, the 1:10 dataset is seen to increase at a lower rate than 1:6. This is in contrast to the findings of the Ir complex; however, comparing percentage increases, it is clear that the differences are very small. Therefore, in an experiment, there is no real motivation to use one over the other. This observation of the changes to the Rh complex unit cell again demonstrates that the long-range damage timescales reach a maximum close to or shortly after the 1:6 X-ray free gap of 12~s (3~s $+$ 9~s data saving), and as such, increasing the gap duration to 21~s (12~s $+$ 9~s data saving) will make very little difference to the overall structural damage. In the 1:2 case, however, the diffraction patterns probe early in the radical diffusion lifetime. 

\subsection{Changes to the Molecular Unit}

Rietveld refinements were conducted on both \ce{[Ir(COD)Cl]2} and \ce{[Rh(COD)Cl]2} datasets to probe any possible dependence on irradiation-to-gap duration on the atomic coordinates of the molecular unit, averaged across all unit cells, see Figure~\ref{fig:Ir_Rh_rietveld}. Changes to the lattice parameters, bond angles and interatomic distances of the central M--Cl core structure are plotted corresponding to a dose of 89~MGy, relative to the initial undamaged structure, across the three irradiation regimes. This dose was chosen as it corresponds to the maximum dose absorbed by the Rh complex after a 2~hour experiment time and enables comparison of structural changes with the Ir complex at the same absorbed dose.

\begin{figure*}[h!]
\centering
\includegraphics[width=\textwidth]{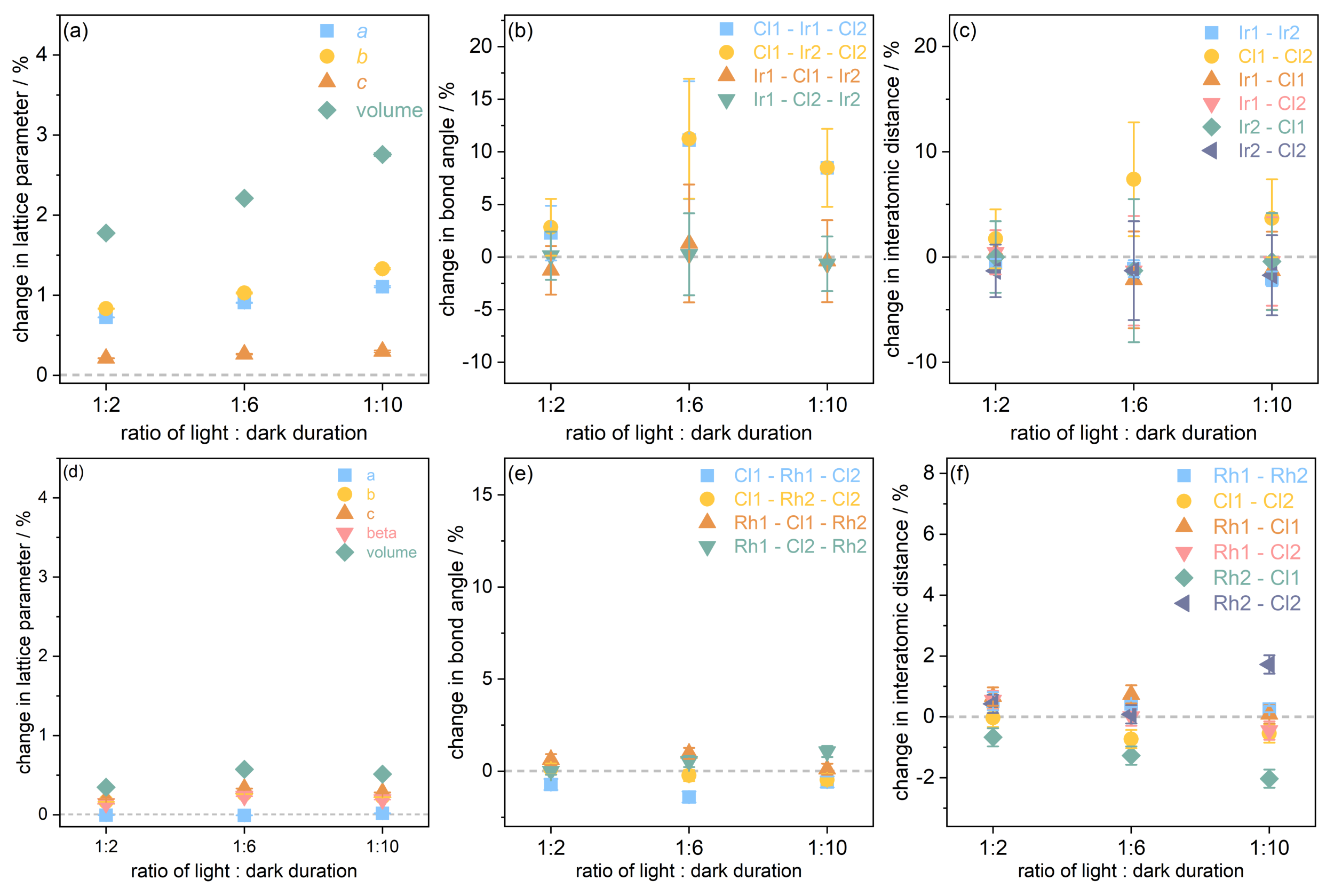}
\caption{\label{fig:Ir_Rh_rietveld}Quantitative results obtained from Rietveld refinements of (a)-(c) \ce{[Ir(COD)Cl]2} and (d)-(f) \ce{[Rh(COD)Cl]2}, at 89.0~MGy and 34.8~MGy, respectively. The plots present the change in (a) and (d) lattice parameters, (b) and (e) bond angles and (c) and (f) interatomic distances of the central M--Cl motif.} 
\end{figure*}

Considering first the percentage changes to lattice parameters for both complexes, Figure~\ref{fig:Ir_Rh_rietveld}(a) and (d), obtained from the Rietveld refinements, it is evident that the trends in lattice parameters increase across the three irradiation regimes, agree with those obtained from Le Bail refinements. Correlating the changes to bond angles and interatomic distances extracted from the structure files initially for the Ir complex, Figure~\ref{fig:Ir_Rh_rietveld}(b) and (c), the Cl--Ir--Cl angles increase while the Ir--Cl--Ir overall a experience decrease in bond angle to a lesser extent (although at 1:6 this is seen to increase very slightly but lies within the margins of error). Trends in interatomic distances are more subtle, however, the Cl--Cl distance undergoes a substantial expansion relative to the other atomic pairs, which is particularly pronounced in the 1:6 case ($+$7.2~\%). This consistent Cl--Cl expansion is in agreement with the results previously reported by the authors of this work,~\cite{Fernando2021} wherein the motility of Cl atoms within the M--Cl core was attributed to the loss of Cl observed in XPS. The initial (minimum XRD dose) interatomic distances and bond angles and the corresponding values at a dose of 89~MGy and 34.8~MGy for the Ir and Rh complex, respectively, are captured in Tables S2 and S3 in the Supplementary Information.\par

In the 1:2 case, the Cl1 swings back from its initial bent structure about the Ir--Ir axis and extends slightly out relative to the M--Cl core. This explains the increase in Cl--Ir--Cl angle observed ($+$3~\%) and the very slight increase in Cl1--Cl2 ($+$1.8~\%) due to the Cl1 pushing away from the central M--Cl structure. For the 1:6 structure at 89~MGy, the Cl2 atom, instead of the Cl1 atom, is found to swing back drastically about the Ir--Ir axis such that there is a flattening of the central M--Cl about the M--M axis. Increasing the duration of the gap to 12~s sees the Cl--Ir--Cl angle increase about the M--M axis by approximately 5~\%. This widening of the central motif about the M--M axis agrees with the previous characterisation of PXRD radiation damage in the Ir complex.~\cite{Fernando2021} The order of Cl--Ir--Cl angle increase from smallest to largest is 1:2, followed by 1:10, then 1:6. From Le Bail refinements, it is expected that 1:6 has a smaller angular increase than the 1:10 structure. There is no significant difference between Ir--Cl distances across the three irradiation regimes. Overall, Rietveld refinements show consistently that the M--Cl rhombus flattens predominantly about the Ir--Ir axis. This is mainly attributed to the movement of Cl atoms (either Cl1 or Cl2). The Ir atoms move slightly, but the overall change in the central part of the molecular unit is dominated by the movement of Cl atoms. These changes to atomic coordinates of the M--Cl core can be seen in Figure~\ref{fig:Ir_rietstructure}, which presents a heat map of the atomic displacement magnitudes of the 89~MGy structure relative to the initial, minimum dose structure. \par 

\begin{figure*}[h!]
\centering
\includegraphics[width=0.7\textwidth]{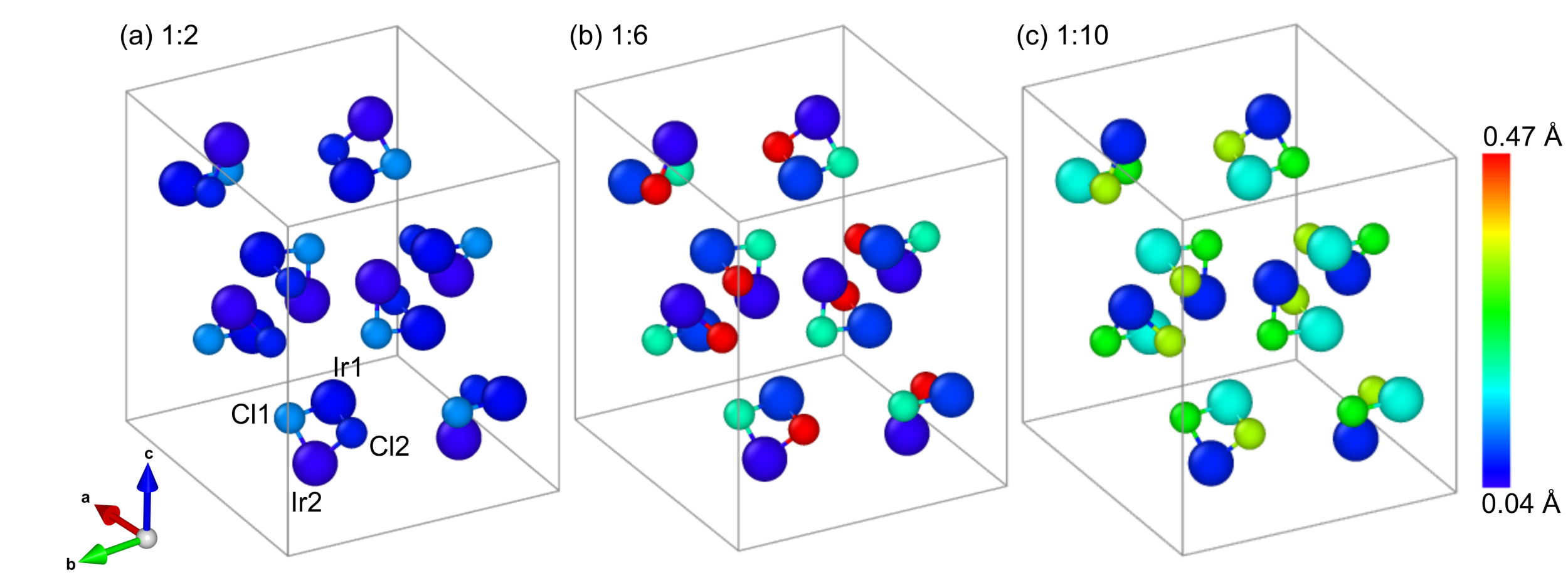}
\caption{\label{fig:Ir_rietstructure}The magnitude of atomic displacements, in \AA, of the central M--Cl core of the \ce{[Ir(COD)Cl]2} unit cell after an absorbed dose of 89~MGy, relative to the minimum dose structure for the three regimes of X-ray irradiation to X-ray-free ratios of (a) 1:2, (b) 1:6 and (c) 1:10. Figure produced using the Ovito software package.~\cite{ovito2010}} 
\end{figure*}

In the case of \ce{[Rh(COD)Cl]2}, all atomic displacements are too subtle to observe visually (see Figure~S4 in the Supplementary Information). However, the quantitative data extracted from the Rietveld refinement, see Figure~\ref{fig:Ir_Rh_rietveld}(e) and (f), show a reduction in Cl--Cl distances for all irradiation regimes and an increase in Rh--Rh distances (opposite to Ir complex), as was observed in the authors' previous work.~\cite{Fernando2022} There is a marginal increase in Rh--Cl--Rh angle with a marginal decrease in Cl--Rh--Cl angle. The small changes to atomic coordinates of the Rh--Cl core are evident in the heat map of atomic displacement magnitudes of the 34.8~MGy structure relative to the initial, minimum dose structure. Results of the Rietveld refinement for the Rh complex show agreement with the Le Bail refinements in the trend that 1:6 undergoes the greatest structural change; however, the extent of these changes is minimal, and no definitive conclusions can be drawn. \par 

To summarise, moving from the 1:2 irradiation regime, where a 3~s wait X-ray-on period follows a diffraction pattern collection, to the 1:6 regime, where the 3~s wait period is dark, systematic structural differences are observed. These are evidenced by a systematic increase in lattice parameters, Cl--M--Cl bonds, as well as an expansion of the Cl--Cl bond. Thus, the impact of reducing the light:dark ratio from 1:2 to 1:6 on the dark progression of long-scale damage is clear. The long-range cascade of radicals appears to continue past the timescales of the 1:2 data collection. It is interesting to note that increasing the duration of the dark time from 12~s (1:6) to 21~s (1:10), sees a drop in Cl--M--Cl bond and Cl--Cl increase. This is despite all lattice parameters continuing to increase.\par 

Previous UHV XPS experiments showed that the molecular unit undergoes substantial local loss in Cl from the onset of irradiation, even at relatively low X-ray doses.~\cite{Fernando2021} A potential explanation for the behaviour observed in the present PXRD results could be that Cl loss continues to occur during the dark period. The local cleaving of the Cl--M bonds and the formation of Cl species outside the main molecular unit, could mean that of the eight molecular units in the Ir complex and four in the Rh complex, after 21~s of X-ray free time, there are some molecules which are deficient of a Cl atom and others which are not. As such, the averaged atomic positions extracted from Rietveld refinements may no longer be robust enough to describe reliably the M--Cl coordination. An explanation of why this is not observed in the 1:6 dataset, despite also losing Cl atoms, could be that there is a threshold after which the majority of the Cl atoms in the irradiated sample are lost, and therefore the information obtained from Rietveld refinements no longer describes the majority of molecular units. However at 1:6, despite some loss in Cl, the majority of Cl atoms could still be bound to the metal atoms.\par

In addition, temperature effects can also play a role. The temperature change of the sample owing to the X-ray beam may peak before it gradually dissipates and cools, in the absence of further irradiation during the dark period. It is, therefore, expected that the sample temperature after 12~s (1:6) of dark time is greater than after a 21~s gap. As such, long-range heat dissipation (and sample temperature) may play a greater role than photoionisation effects in dark periods, suggesting that the atomic coordinate and bond differences between the 1:6 and 1:10 datasets could be due to thermal atomic vibrations. However, crucially, this temperature hypothesis does not explain why the 1:2 regime experiences smaller structural change relative to the 1:6 regime, so is highly unlikely to be applicable in this experiment. 

For \ce{[Ir(COD)Cl]2}, the systematic increase in lattice parameters from 1:2 to 1:6 to 1:10 irradiation regimes, despite the apparent easing of Cl displacement, can be explained by the overall increase in disorder (thermally-driven or otherwise) and the possible radiation-induced formation and build-up of \ce{H2}, resulting in growing lattice strain.

\subsection{Changes to Chemical Environment}

\begin{figure*}[h!]
\centering
\includegraphics[width=0.7\textwidth]{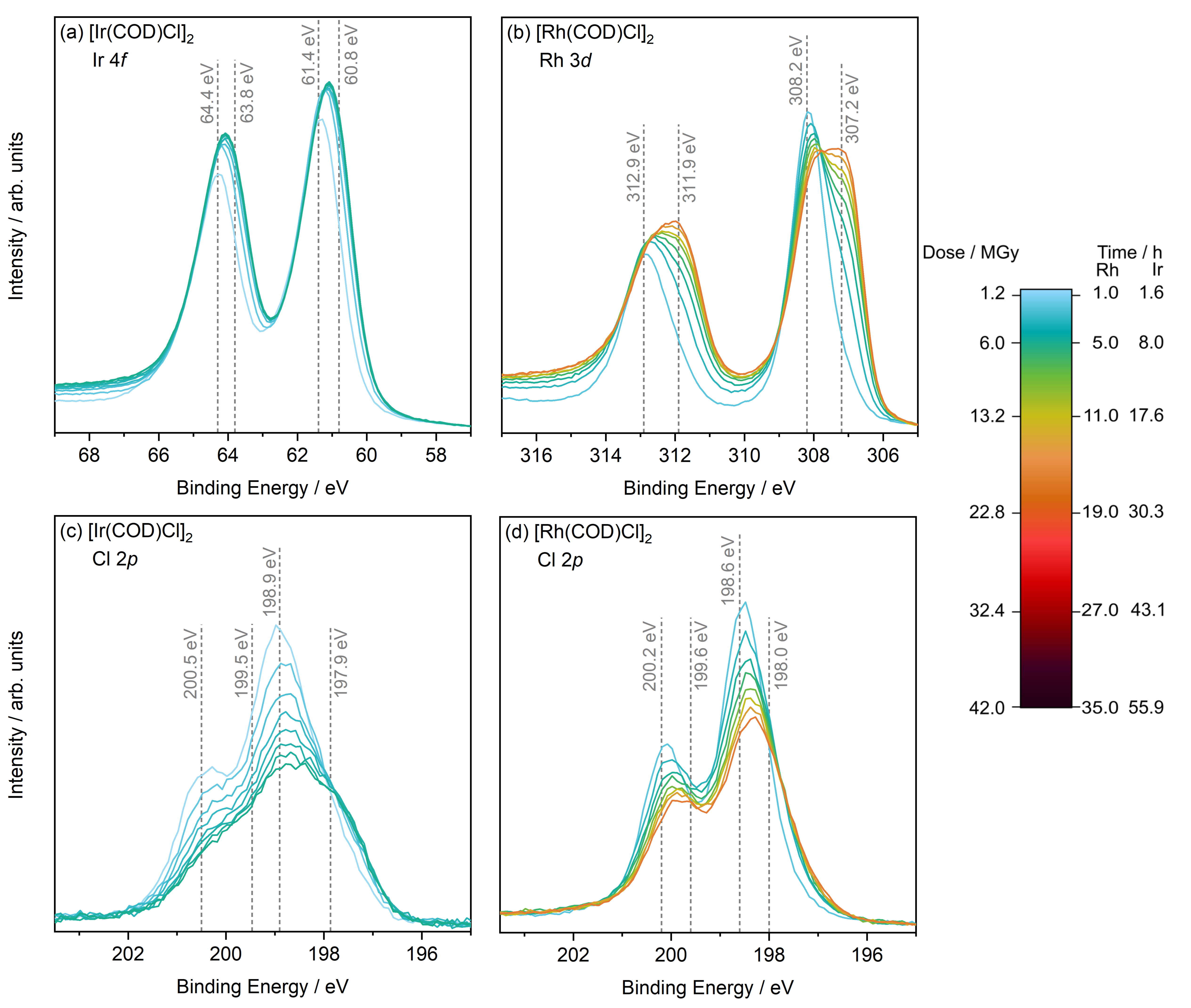}
\caption{\label{fig:XPS_iterations_gap}Averaged core level X-ray photoelectron spectra of the \ce{[M(COD)Cl]2} catalysts as a function of X-ray exposure during gap measurements, including (a) Ir 4\textit{f} and (b) Rh 3\textit{d} spectra of \ce{[Ir(COD)Cl]2} and \ce{[Rh(COD)Cl]2}, respectively, and (c) and (d) Cl 2\textit{p} spectra. The legend includes the measurement time as well as the calculated X-ray dose as Average Dose Whole Crystal (AD-WC) from RADDOSE-3D. The grey dotted lines and binding energy values shown correspond to the binding energy positions of the main spectral contributions present in the first spectra (2~h).}
\end{figure*}

To determine whether there is an influence of X-ray-free dark periods on the specific local chemical changes, a complementary XPS study was carried out. The main metal Ir~4\textit{f} and Rh~3\textit{d} core levels, were collected along with Cl~2\textit{p}, according to the steps outlined in the Methods section above. From Figure~\ref{fig:XPS_iterations_gap}(a), it is evident that even when introducing dark periods during measurement, both metal peaks show growing contributions at lower binding energies, over the course of 16~hours, signalling a reduction from their initial $+$1 state. In addition, the Cl~2\textit{p} spectra show a drastic loss in intensity for both complexes, see Figures~\ref{fig:XPS_iterations_gap}(b) and (d), with increasing cumulative X-ray dose up to a maximum dose of 11~MGy and 19~MGy for the Ir and Rh complex, respectively. These observations are in agreement with the previous continuous exposure results reported.~\cite{Fernando2021} Correlating these observations with the structural changes observed in PXRD, at the maximum XPS dose of 11 and 19~MGy, PXRD refinements show that the samples have already experienced some unit cell expansion. As such, it is understood that in both continuous and gap collection methods, the local chemical changes occur concurrently with the overall structural (unit cell) change at this dose. As in the continuous case, at the cumulative X-ray doses probed in this experiment, the C~1\textit{s} spectra for both metal complexes underwent very little to no chemical environment change, see Figure~S5 in the Supplementary Information.\par

Peak fit analysis enabled the quantification of local changes that occur when sample irradiation is interrupted by the momentary movement of the beam away from the measurement spot. These included the metal photoreduction, loss of Cl, the growing presence of a new Cl species, plotted as a function of X-ray dose and X-ray exposure time, as well as the shifting of the valence band maximum (VBM) towards the Fermi energy E\textsubscript{F}, see Figure~\ref{fig:XPS_gap_cont}. The Ir and Rh photoreduction, presented in Figure~\ref{fig:XPS_gap_cont}(a) and (e), respectively, stabilises at a certain absorbed dose of 5~MGy, which coincides with the onset of stabilisation of Cl loss in Figures~\ref{fig:XPS_gap_cont}(b) and (f). Considering the shift of the VBM towards the E\textsubscript{F}, shown in Figure~\ref{fig:XPS_gap_cont}(d) and (h), it is evident that after 4 and 3--5~MGy, the VBM to E\textsubscript{F} distances of both Ir and Rh, approaches zero, signalling that both systems become metallic under these experimental conditions.\par  

\begin{figure*}[h!]
\centering
\includegraphics[width=0.92\linewidth]{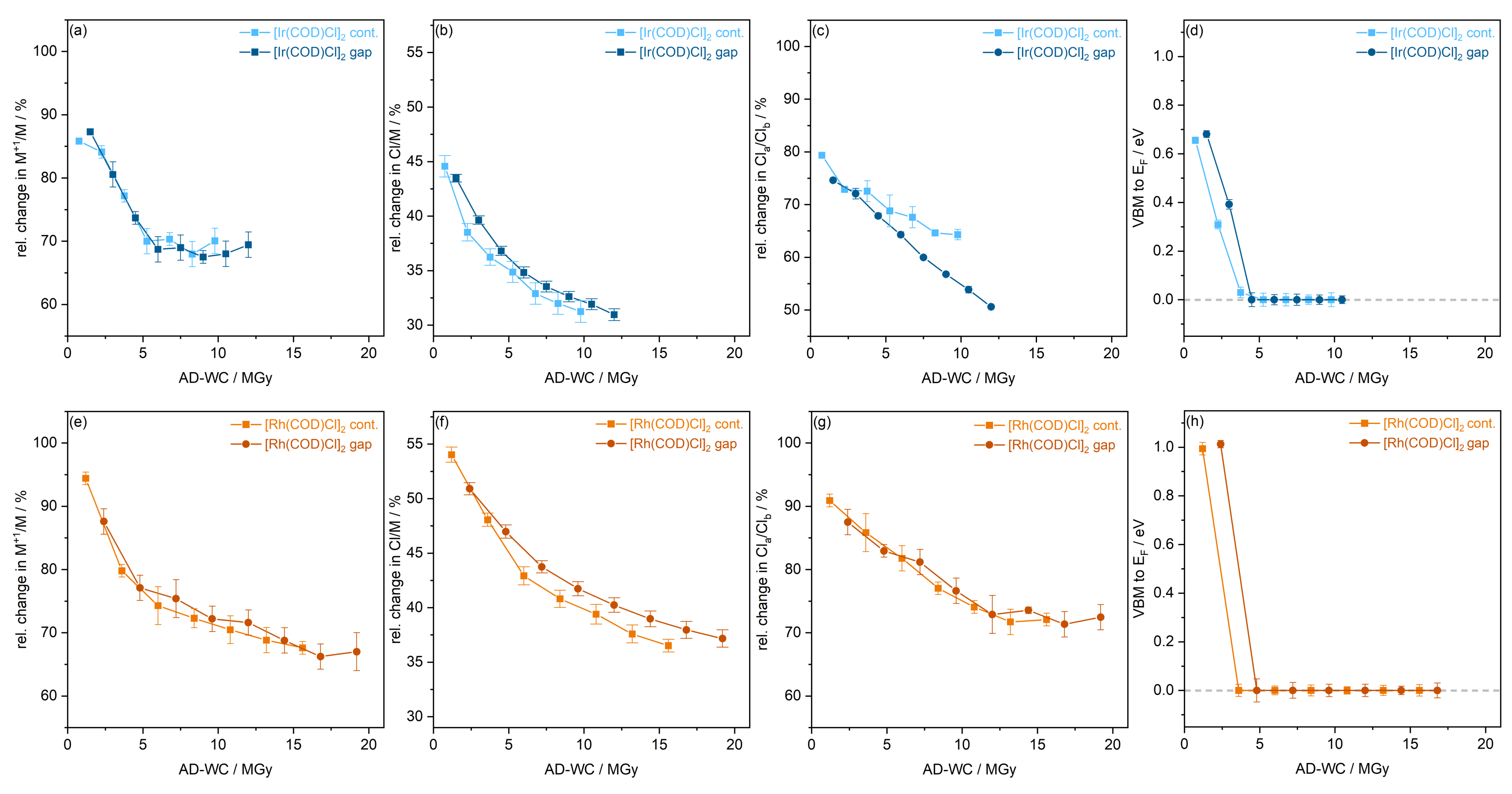}
\caption{\label{fig:XPS_gap_cont}Comparative plots of the quantitative changes to photoelectron spectra as a function of dose and X-ray exposure time extracted from core level and valence state analysis, between the continuous and gap irradiation regimes. (a)-(d) \ce{[Ir(COD)Cl]2} and (e)-(h) \ce{[Rh(COD)Cl]2}. (a) and (e) Relative atomic percentages (RAPs) (rel.~at.~\% in plot) of the ratios of the Ir and Rh core level contributions. (b) and (f) RAP ratios of the total M and Cl signals (c) and (g) RAPs of the ratios of the two Cl environments observed, where Cl\textsubscript{a} is the main Cl environment and Cl\textsubscript{b} is the additional, lower BE feature. (d) and (h) Distance between VBM and $E_F$.}
\end{figure*}

These chemical changes are then compared with those observed when the measurement point is irradiated continuously, with spectra collected every two hours. Upon direct visual comparison of the core level spectra of the complexes irradiated continuously, with those obtained with gaps incorporated into the measurement, with respect to the absorbed dose only subtle differences are observed (see Figure~S6 in the Supplementary Information). At an approximately equal given dose of 7--8~MGy for the Ir complex, the continuous and gap Ir~4\textit{f} spectra show no notable difference. The Cl~2\textit{p} spectra are very similar, although there is a very slight increase in intensity at the lower binding energy component for the gap spectra. At 6--7~MGy, the continuous and gap Rh~3\textit{d} and Cl~2\textit{p} spectra of the Rh complex are identical within the accuracy of the experiments.\par  

Quantitative analysis of the spectra further underlines these qualitative observations. Figure~\ref{fig:XPS_gap_cont}(a) shows that in both Ir and Rh complexes, the reduction in the metal centre is comparable across the two irradiation regimes within the experimental errors. Both setups experienced a $\approx$30\% drop of Ir\textsuperscript{1+} and 20\% drop of Rh\textsuperscript{1+} for the two complexes, respectively. Similarly, the overall loss of Cl relative to the metal which sees a 15\% loss under both irradiation regimes, Figures~\ref{fig:XPS_gap_cont}(b) and (f).\par

The relative changes in the two Cl species present in the samples are also explored, where Cl\textsubscript{a} corresponds to the intrinsic Cl environment and Cl\textsubscript{b}, the formation of a new Cl species, see Figure~\ref{fig:XPS_gap_cont}(c) and (g). In the Ir complex, there is initially no notable difference in Cl\textsubscript{a}/Cl\textsubscript{b} between the irradiation regimes. However, after a dose of 7~MGy, the proportion of Cl\textsubscript{a} in the continuously irradiated sample stabilises whilst, in the gap measurement, the Cl\textsubscript{b} continues to decrease linearly. This divergence at higher dose is likely due to the substantial decrease in Cl~2\textit{p} resolution as Cl is liberated from the complex, which introduces greater uncertainties in the results of the peak fit. For the Rh complex, see Figure~\ref{fig:XPS_gap_cont}(g), it is evident that there is no difference in Cl\textsubscript{a}/Cl\textsubscript{b} trends when continuously irradiating and introducing X-ray free gaps. The VBM to E\textsubscript{F} shift (Figures~\ref{fig:XPS_gap_cont}(d) and (h)) again show no notable difference between the irradiation setups. VB spectra show that under the two irradiation regimes, both systems become metallic after approximately 4 and 3--5~MGy dose for Ir and Rh, respectively.\par

As no notable difference in local chemical change can be observed for both complexes when periodic dark periods are introduced in the XPS measurements, it can be deduced that no dark progression occurs. This observation confirms that timescales of such site-specific events, e.g. metal reduction (which is well understood to be of the femtosecond order) and Cl loss, are very short relative to the irradiation period in this experiment. As such, the local damage processes have essentially finished within the duration of the irradiation pulse, and no further damage occurs in the dark period.

\section{Conclusions}

Experiments were carried out to determine the influence of introducing X-ray-free dark periods during X-ray characterisation on both global damage and specific damage via PXRD and XPS, respectively. Profile and structural refinements of the equivalent data for both \ce{[Ir(COD)Cl]2} and \ce{[Rh(COD)Cl]2} showed conclusively that the 1:2 irradiation regime experienced the least global damage, using metrics of lattice parameter expansion and atomic displacements. The 1:6 and 1:10 irradiation regimes showed the greatest global damage, with the 1:6 showing the greatest movement of Cl in the central M--Cl structure. This can be attributed to the timescales of global damage progression in the sample. In the setting with the smallest proportion of dark time (1:2), it is hypothesised that the damage progression is cut off before it is complete, resulting in a relatively small overall global change. With long dark periods, the dark damage progresses further and for longer, resulting in greater structural change when the diffraction pattern is next collected. An explanation proposed is that dark progression does indeed continue in the 1:10 regime, resulting in a greater number of Cl atoms being liberated from each molecular unit, which is not taken into account in the averaged structures obtained from Rietveld refinements.\par
To determine whether site-specific dark damage can be observed in these systems, XPS experiments were conducted, comparing a continuous X-ray irradiation regime with one where short dark periods were incorporated into the measurement. Comparing the local chemical changes between these two regimes showed no significant difference in the extent of sample damage. This indicates that the local photoreduction and loss in Cl are short-lived processes and that these timescales are much shorter than the duration of the probe X-ray pulse so that all possible damage has occurred before the dark period begins.\par
These studies confirm the importance of reducing, as much as possible, X-ray-free dark periods during measurements, which in X-ray experiments typically originate from slow detector read-out times to outrun global damage. This is in agreement with the work on protein crystals by Owen~\textit{et al.}~\cite{Owen2012} Stopping exposure to X-rays momentarily does not, as might be expected, relax the system (and reduce damage) as dark progression is shown here to occur even on second timescales during synchrotron experiments.\par 

Crucially, through the systematic light:dark experimental approaches outlined in this work, the occurrence of global dark progression on these model catalysts was confirmed. These results strongly motivate future computational studies--- to model the differences in kinetic and chemical processes under continuous and discontinuous irradiation, with varying X-ray-free gaps and shed further light on the damage pathways.

\section*{Data Availability}
Data for this article, including outputs of XRD refinements and XPS peak fit analysis, are available at Zenodo in Origin format at https://doi.org/10.5281/zenodo.13959906.

\section*{Conflicts of interest}
There are no conflicts to declare.

\section*{Acknowledgements}
NKF acknowledges support from the Engineering and Physical Sciences Research Council (EP/L015277/1). This work was carried out with the support of Diamond Light Source, instrument I11 (proposal EE18324-1).

%%%REFERENCES%%%
\bibliography{ref} 

\begin{thebibliography}{1}

\bibitem{Fernando2021}
N.~K. Fernando, A.~B. Cairns, C.~A. Murray, A.~L. Thompson, J.~L. Dickerson,
  E.~F. Garman, N.~Ahmed, L.~E. Ratcliff, and A.~Regoutz.
\newblock {Structural and Electronic Effects of X-ray Irradiation on
  Prototypical [M(COD)Cl]$_2$ Catalysts}.
\newblock {\em The Journal of Physical Chemistry A}, 125(34):7473--7488, 9
  2021.

\bibitem{ovito2010}
A.~Stukowski.
\newblock {Visualization and analysis of atomistic simulation data with
  OVITO–the Open Visualization Tool}.
\newblock {\em Modelling and Simulation in Materials Science and Engineering},
  18(1):015012, 1 2010.

\end{thebibliography}


\begin{thebibliography}{10}

\bibitem{Blundell1976}
T.~L. Blundell and L.~N. Johnson.
\newblock {\em {Protein Crystallography}}.
\newblock New York, London Academic Press, 1976.

\bibitem{Brooks2017}
J.~C. Brooks-Bartlett, R.~A. Batters, C.~S. Bury, E.~D. Lowe, H.~M. Ginn, A.~Round, and E.~F. Garman.
\newblock {Development of tools to automate quantitative analysis of radiation damage in SAXS experiments}.
\newblock {\em Journal of Synchrotron Radiation}, 24(1):63--72, 1 2017.

\bibitem{Bury2018}
C.~S. Bury, J.~C. Brooks-Bartlett, S.~P. Walsh, and E.~F. Garman.
\newblock {Estimate your dose: RADDOSE-3D}.
\newblock {\em Protein Science}, 27(1):217--228, 1 2018.

\bibitem{Chapman2011}
H.~N. Chapman, P.~Fromme, A.~Barty, T.~A. White, R.~A. Kirian, A.~Aquila, M.~S. Hunter, J.~Schulz, D.~P. DePonte, U.~Weierstall, R.~B. Doak, F.~R. N.~C. Maia, A.~V. Martin, I.~Schlichting, L.~Lomb, N.~Coppola, R.~L. Shoeman, S.~W. Epp, R.~Hartmann, D.~Rolles, A.~Rudenko, L.~Foucar, N.~Kimmel, G.~Weidenspointner, P.~Holl, M.~Liang, M.~Barthelmess, C.~Caleman, S.~Boutet, M.~J. Bogan, J.~Krzywinski, C.~Bostedt, S.~Bajt, L.~Gumprecht, B.~Rudek, B.~Erk, C.~Schmidt, A.~H{\"{o}}mke, C.~Reich, D.~Pietschner, L.~Str{\"{u}}der, G.~Hauser, H.~Gorke, J.~Ullrich, S.~Herrmann, G.~Schaller, F.~Schopper, H.~Soltau, K.-U. K{\"{u}}hnel, M.~Messerschmidt, J.~D. Bozek, S.~P. Hau-Riege, M.~Frank, C.~Y. Hampton, R.~G. Sierra, D.~Starodub, G.~J. Williams, J.~Hajdu, N.~Timneanu, M.~M. Seibert, J.~Andreasson, A.~Rocker, O.~J{\"{o}}nsson, M.~Svenda, S.~Stern, K.~Nass, R.~Andritschke, C.-D. Schr{\"{o}}ter, F.~Krasniqi, M.~Bott, K.~E. Schmidt, X.~Wang, I.~Grotjohann, J.~M. Holton, T.~R.~M. Barends, R.~Neutze, S.~Marchesini, R.~Fromme,
  S.~Schorb, D.~Rupp, M.~Adolph, T.~Gorkhover, I.~Andersson, H.~Hirsemann, G.~Potdevin, H.~Graafsma, B.~Nilsson, and J.~C.~H. Spence.
\newblock {Femtosecond X-ray protein nanocrystallography}.
\newblock {\em Nature}, 470(7332):73--77, 2 2011.

\bibitem{Christensen2019}
J.~Christensen, P.~N. Horton, C.~S. Bury, J.~L. Dickerson, H.~Taberman, E.~F. Garman, and S.~J. Coles.
\newblock {Radiation damage in small-molecule crystallography: fact not fiction}.
\newblock {\em IUCrJ}, 6(4):703--713, 7 2019.

\bibitem{Coelho2018}
A.~A. Coelho.
\newblock {TOPAS and TOPAS-Academic : an optimization program integrating computer algebra and crystallographic objects written in C++}.
\newblock {\em Journal of Applied Crystallography}, 51(1):210--218, 2 2018.

\bibitem{TOPAS7}
A.~A. Coelho.
\newblock {TOPAS Academic: General Profile and Structure Analysis Software for Powder Diffraction Data, version 7.1 (Computer Software), Coelho Software: Brisbane, QLD}, 2020.

\bibitem{Dinnebier2018}
R.~E. Dinnebier, A.~Leineweber, and J.~S.~O. Evans.
\newblock {The Rietveld Method}.
\newblock In {\em Rietveld Refinement Practical Powder Diffraction Pattern Analysis using TOPAS}, chapter~2, pages 16--88. De Gruyter, 2018.

\bibitem{Fernando2022}
N.~K. Fernando, H.~L.~B. Bostr{\"{o}}m, C.~A. Murray, R.~L. Owen, A.~L. Thompson, J.~L. Dickerson, E.~F. Garman, A.~B. Cairns, and A.~Regoutz.
\newblock {Variability in X-ray induced effects in [Rh(COD)Cl] 2 with changing experimental parameters}.
\newblock {\em Physical Chemistry Chemical Physics}, 24(46):28444--28456, 2022.

\bibitem{Fernando2021}
N.~K. Fernando, A.~B. Cairns, C.~A. Murray, A.~L. Thompson, J.~L. Dickerson, E.~F. Garman, N.~Ahmed, L.~E. Ratcliff, and A.~Regoutz.
\newblock {Structural and Electronic Effects of X-ray Irradiation on Prototypical [M(COD)Cl]$_2$ Catalysts}.
\newblock {\em The Journal of Physical Chemistry A}, 125(34):7473--7488, 9 2021.

\bibitem{Dig_Sco_2020}
C.~Kalha, N.~K. Fernando, and A.~Regoutz.
\newblock {Digitisation of Scofield Photoionisation Cross Section Tabulated Data}.
\newblock {\em figshare}, 2020.

\bibitem{Kmetko2006}
J.~Kmetko, N.~S. Husseini, M.~Naides, Y.~Kalinin, and R.~E. Thorne.
\newblock {Quantifying X-ray radiation damage in protein crystals at cryogenic temperatures}.
\newblock {\em Acta Crystallographica Section D Biological Crystallography}, 62(9):1030--1038, 9 2006.

\bibitem{LeBail1988}
A.~Le~Bail, H.~Duroy, and J.~Fourquet.
\newblock {Ab-initio structure determination of LiSbWO6 by X-ray powder diffraction}.
\newblock {\em Materials Research Bulletin}, 23(3):447--452, 3 1988.

\bibitem{Mcgeehan2009}
J.~McGeehan, R.~B.~G. Ravelli, J.~W. Murray, R.~L. Owen, F.~Cipriani, S.~McSweeney, M.~Weik, and E.~F. Garman.
\newblock {Colouring cryo-cooled crystals: online microspectrophotometry}.
\newblock {\em Journal of Synchrotron Radiation}, 16(2):163--172, 3 2009.

\bibitem{Meents2010}
A.~Meents, S.~Gutmann, A.~Wagner, and C.~Schulze-Briese.
\newblock {Origin and temperature dependence of radiation damage in biological samples at cryogenic temperatures}.
\newblock {\em Proceedings of the National Academy of Sciences}, 107(3):1094--1099, 1 2010.

\bibitem{Neutze2000}
R.~Neutze, R.~Wouts, D.~van~der Spoel, E.~Weckert, and J.~Hajdu.
\newblock {Potential for biomolecular imaging with femtosecond X-ray pulses}.
\newblock {\em Nature}, 406(6797):752--757, 8 2000.

\bibitem{Owen2012}
R.~L. Owen, D.~Axford, J.~E. Nettleship, R.~J. Owens, J.~I. Robinson, A.~W. Morgan, A.~S. Dor{\'{e}}, G.~Lebon, C.~G. Tate, E.~E. Fry, J.~Ren, D.~I. Stuart, and G.~Evans.
\newblock {Outrunning free radicals in room-temperature macromolecular crystallography}.
\newblock {\em Acta Crystallographica Section D Biological Crystallography}, 68(7):810--818, 7 2012.

\bibitem{Ravelli2000}
R.~B. Ravelli and S.~M. McSweeney.
\newblock {The ‘fingerprint’ that X-rays can leave on structures}.
\newblock {\em Structure}, 8(3):315--328, 3 2000.

\bibitem{Rietveld1967}
H.~M. Rietveld.
\newblock {Line profiles of neutron powder-diffraction peaks for structure refinement}.
\newblock {\em Acta Crystallographica}, 22(1):151--152, 1 1967.

\bibitem{Scofield_1973}
J.~Scofield.
\newblock {Theoretical photoionization cross sections from 1 to 1500 keV.}
\newblock {\em Technical Report; Lawrence Livermore Laboratory}, pages UCRL--51326, 1 1973.

\bibitem{Davies2007}
R.~J. Southworth-Davies, M.~A. Medina, I.~Carmichael, and E.~F. Garman.
\newblock {Observation of Decreased Radiation Damage at Higher Dose Rates in Room Temperature Protein Crystallography}.
\newblock {\em Structure}, 15(12):1531--1541, 12 2007.

\bibitem{ovito2010}
A.~Stukowski.
\newblock {Visualization and analysis of atomistic simulation data with OVITO–the Open Visualization Tool}.
\newblock {\em Modelling and Simulation in Materials Science and Engineering}, 18(1):015012, 1 2010.

\bibitem{Thompson2009}
S.~P. Thompson, J.~E. Parker, J.~Potter, T.~P. Hill, A.~Birt, T.~M. Cobb, F.~Yuan, and C.~C. Tang.
\newblock {Beamline I11 at Diamond: A new instrument for high resolution powder diffraction}.
\newblock {\em Review of Scientific Instruments}, 80(7):075107, 7 2009.

\bibitem{Warkentin2011}
M.~Warkentin, R.~Badeau, J.~Hopkins, and R.~E. Thorne.
\newblock {Dark progression reveals slow timescales for radiation damage between T = 180 and 240 K}.
\newblock {\em Acta Crystallographica Section D Biological Crystallography}, 67(9):792--803, 9 2011.

\bibitem{Warkentin2012}
M.~Warkentin, R.~Badeau, J.~B. Hopkins, A.~M. Mulichak, L.~J. Keefe, and R.~E. Thorne.
\newblock {Global radiation damage at 300 and 260 K with dose rates approaching 1 MGy s-1}.
\newblock {\em Acta Crystallographica Section D Biological Crystallography}, 68(2):124--133, 2 2012.

\bibitem{Watts2019}
J.~F. Watts and J.~Wolstenholme.
\newblock {The Electron Spectrum}.
\newblock In {\em An Introduction to Surface Analysis by XPS and AES}, chapter~3, page~93. John Wiley {\&} Sons, 2 edition, 2019.

\bibitem{Zeldin2013}
O.~B. Zeldin, M.~Gerstel, and E.~F. Garman.
\newblock {RADDOSE-3D : time- and space-resolved modelling of dose in macromolecular crystallography}.
\newblock {\em Journal of Applied Crystallography}, 46(4):1225--1230, 8 2013.

\end{thebibliography}
\bibliographystyle{abbrv} 

\end{document}

\centering\noindent \Large{\textbf{{Investigating Discontinuous X-ray Irradiation as a Damage Mitigation Strategy for \ce{[M(COD)Cl]2} Catalysts}\\
Supplementary Information}}\\

\vspace{7.5mm}

\noindent\large{Nathalie K. Fernando,$^{\ast}$\textit{$^{a}$} Claire A. Murray,\textit{$^{b}$} Amber L. Thompson,\textit{$^{c}$} Katherine Milton,\textit{$^{a}$} Andrew B. Cairns,\textit{$^{d}$} and Anna Regoutz\textit{$^{a,e}$}}  \\%Author names go here instead of "Full name", etc.

\vspace{5mm}
\noindent\small\textit{{$^{a}$~Department of Chemistry, University College London, 20 Gordon Street, London, WC1H~0AJ, UK.}\\
\textit{$^{b}$~Diamond Light Source Ltd, Diamond House, Harwell Science and Innovation Campus, Didcot, Oxfordshire, OX11~0DE, UK.}\\
\textit{$^{c}$~Chemical Crystallography, Chemistry Research Laboratory, University of Oxford, South Parks Road, Oxford OX1~3QR, UK.}\\
\textit{$^{d}$~Department of Materials, Imperial College London, Royal School of Mines, Exhibition Road, SW7~2AZ, UK.}\\
\textit{$^{e}$~Present address: Department of Chemistry, Inorganic Chemistry Laboratory, South Parks Road,OX1~3QR, UK.}}\\

\vspace{10mm}

\captionsetup[table]{labelfont=bf,textfont=normalfont}

\begin{table*}[!ht]
    \caption{The parameters and corresponding values included in the RADDOSE-3D input file for \ce{[Ir(COD)Cl]2} and\ce{[Rh(COD)Cl]2} used to estimate the doses in the powder XRD experiments at beamline I11 at Diamond Light Source, Didcot, UK. The same values were used in the studies presented in Fernando~\textit{et al.}\ 2021.~\cite{Fernando2021}}\label{tab:raddose}
    \centering
    \begin{tabular*}{0.7\textwidth}{lll}
        \hline \hline
        RADDOSE parameter &  Beamline I11, DLS (PXRD) &  K-Alpha$+$ (XPS) \\
        \hline
        Crystal type & cylinder & cuboid \\
        Crystal Dimension /$\mu$m & 300 $\times$ 40000 & 60 $\times$ 60 $\times$ 60 \\
        PixelsPerMicron & 0.06 & 0.6\\
        Container material type & mixture & N/A\\
        Material mixture & pyrex & N/A\\
        Container thickness / $\mu$m & 10 & N/A\\
        Container density /   g cm\textsuperscript{-3} & 2.23 & N/A\\
        Beam type & Gaussian & Gaussian\\
        Photon flux / photons/s & 1.5 $\times$ 10\textsuperscript{14}& 3.8 $\times$ 10\textsuperscript{10}\\
        FWHM /$\mu$m & 2000 $\times$ 600 & 326$\times$579\\
        Energy / keV & 15 & 1.487\\
        Collimation type & rectangular & circular \\
        Collimation dimensions / $\mu$m\textsuperscript{2} & 2500 $\times$ 800 & 300 $\times$ 300\\
        \hline \hline
    \end{tabular*}
\end{table*}

\newpage

\begin{figure*}[hb]
\centering
\includegraphics[width=0.8\textwidth]{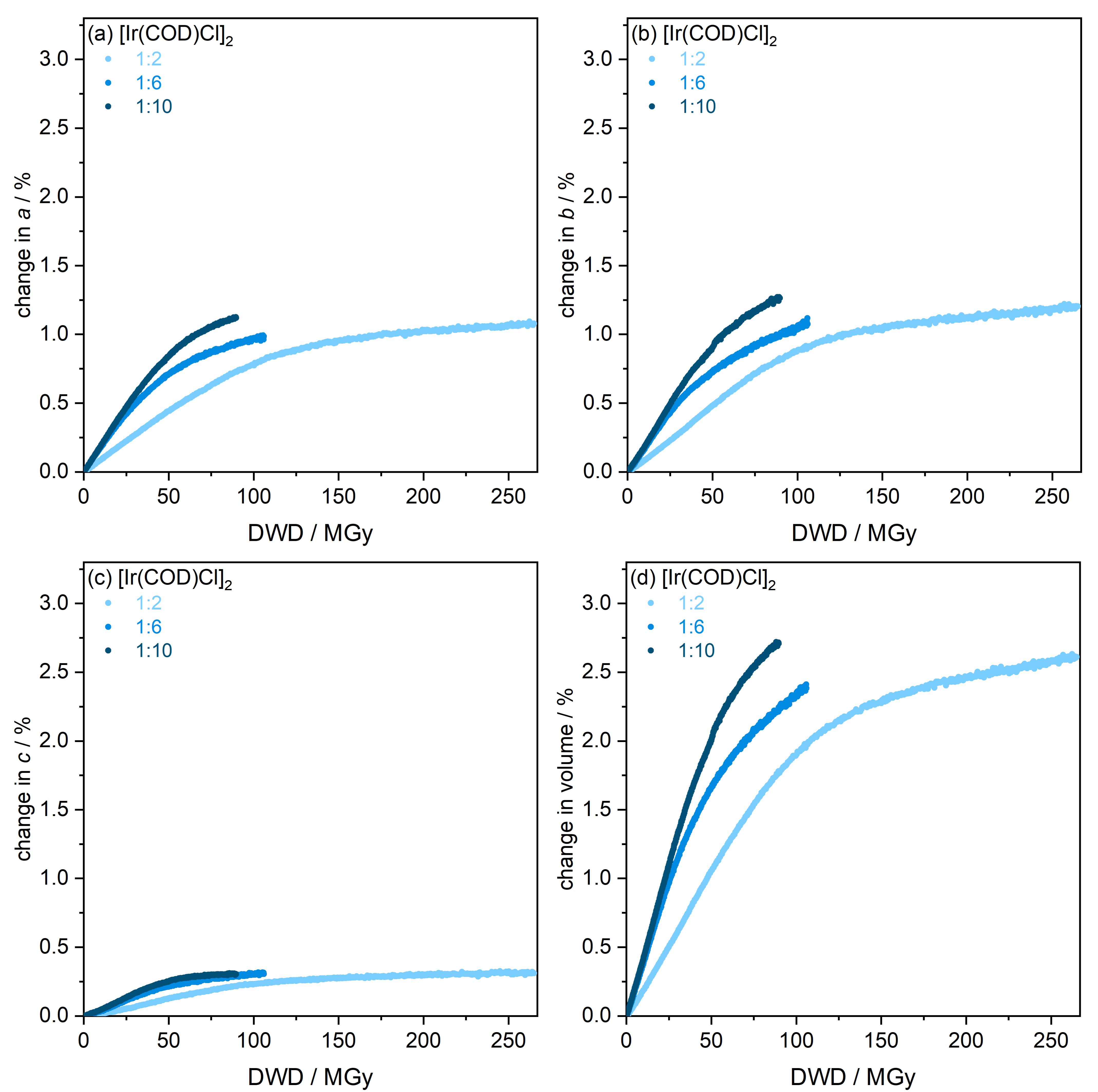}
\caption{\label{fig:Ir_lattice_params} Percentage changes to lattice parameters and unit cell volume, with X-ray dose, for \ce{[Ir(COD)Cl]2}, extracted from Le Bail refinements, for three X-ray irradiation to dark ratios for unit cell axes (a) $a$, (b) $b$, (c) $c$ and (d) unit cell volume. From error propagation calculations, the approximate errors in (a)-(c) are $\pm$0.007\%, $\pm$0.03\% and $\pm$0.007\% for the 1:2, 1:6 and 1:10 datasets, respectively. Similarly, the errors for the change in volume, (d), are $\pm$0.01\%, $\pm$0.06\% and $\pm$0.02\%.}
\end{figure*}

\begin{figure*}[ht]
\centering
\includegraphics[width=\textwidth]{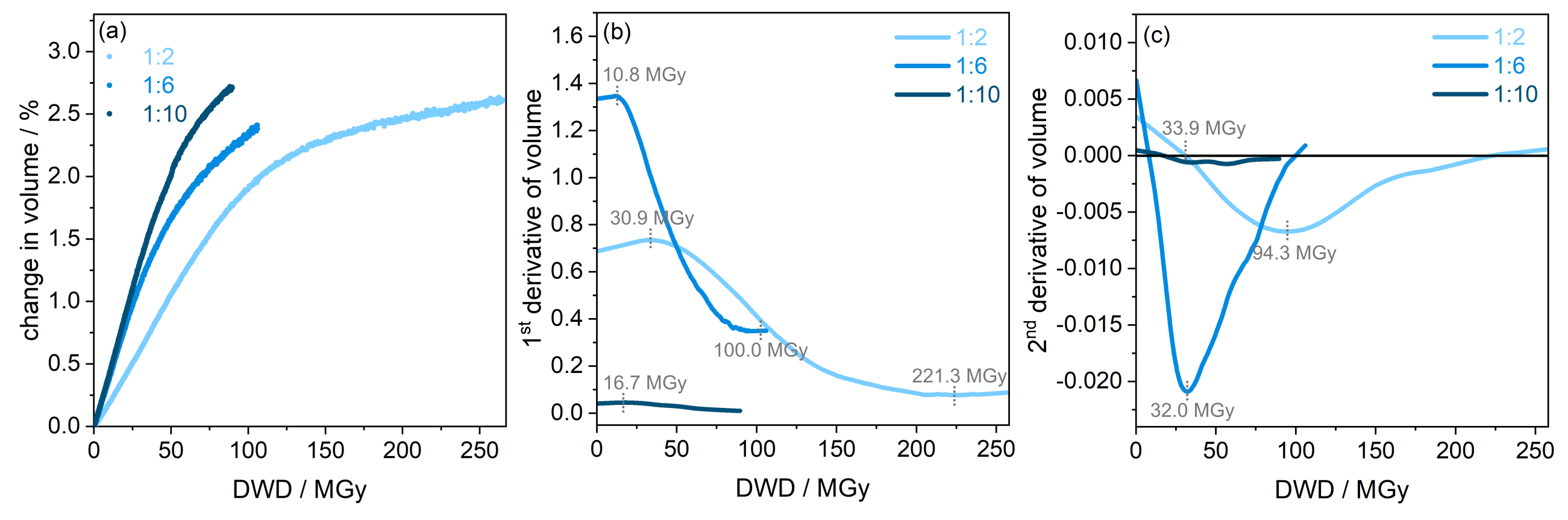}
\caption{\label{fig:Ir_vol_derivatives}The changes in the curvature of the \ce{[Ir(COD)Cl]2} unit cell volume for the 1:2, 1:6 and 1:10 X-ray irradiation/gap regimes as a function of X-ray dose to determine the onset of volume expansion saturation. (a) The percentage change in unit cell volume, (b) the first derivative curve of the volume with maximum and minimum points labelled, and (c) the second derivative of unit cell volume with points of inflexion labelled.}
\end{figure*}

\begin{figure*}[b]
\centering
\includegraphics[width=0.9\textwidth]{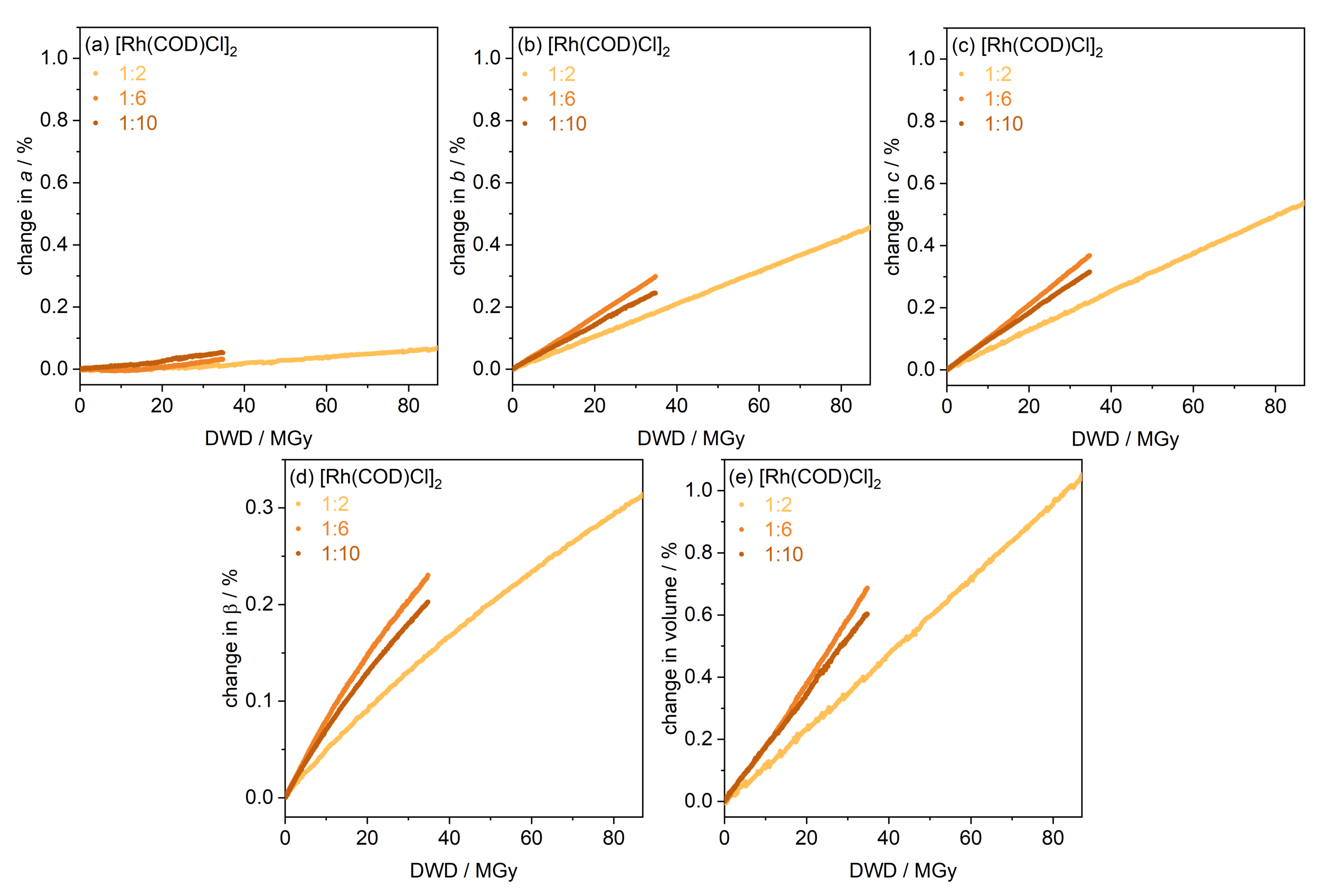}
\caption{\label{fig:Rh_lattice_params}Percentage changes to lattice parameters for \ce{[Rh(COD)Cl]2}, extracted from Le Bail refinements of the PXRD data as a function of X-ray dose and X-ray irradiation to dark ratios for unit cell axes (a) $a$, (b) $b$, (c) $c$, (d) $\beta$ angle and (e) unit cell volume. From error propagation calculations of the last data point, the approximate errors in (a)-(c) are $\pm$0.004\%, $\pm$0.001\% and $\pm$0.002\% for the 1:2, 1:6 and 1:10 datasets, respectively. Similarly, the errors for the change in $\beta$ (d), are $\pm$0.001\% (for all 3 ratios) and for the change in volume, (e), these are $\pm$0.01\%, $\pm$0.003\% and $\pm$0.003\%.}
\end{figure*}

\captionsetup[table]{labelfont=bf,textfont=normalfont}

\begin{table*}[!ht]
    \caption{The interatomic distances and bond angles of \ce{[Ir(COD)Cl]2} at the start of the XRD experiment and at a dose of 89~MGy, for the 1:2, 1:6 and 1:10 light:dark X-ray radiation regimes.}\label{tab:Irbonds}
    \centering
    \begin{tabular*}{0.75\textwidth}{l|c|ll|ll|ll}
        \hline \hline
       & \multirow{2}{*}{Bond} & \multicolumn{2}{c|}{1:2} &  \multicolumn{2}{c|}{1:6} & \multicolumn{2}{c}{1:10} \\
       & & start & end  & start & end  & start & end \\
        \hline
       \multirow{6}{*}{Length(\AA)} & Ir1--Ir2 & 2.927(6) & 2.91(1) & 2.892(8) & 2.86(2) & 2.900(4) & 2.84(2) \\
       & Cl1--Cl2 & 2.89(3) & 2.94(5) & 2.98(4) & 3.2(1) & 2.99(2) & 3.1(1) \\
       & Ir1--Cl1 & 2.28(2)	& 2.29(3) & 2.31(3) & 2.26(7) & 2.32(2) & 2.29(7)\\
       & Ir1--Cl2 & 2.25(2)	& 2.26(3) & 2.29(3) & 2.26(8) & 2.32(2) & 2.31(8)\\
       & Ir2--Cl1 & 2.33(3) & 2.33(4) & 2.32(4) & 2.29(9) & 2.32(2) & 2.31(9)\\
       & Ir2--Cl2 & 2.29(2)	& 2.26(4) & 2.31(3) & 2.28(7) & 2.31(2)	& 2.27(7)\\
        \hline
       \multirow{4}{*}{Angle ($^{\circ}$)} & Cl1--Ir1--Cl2 & 79.2(9) & 81(1) & 81(1) & 90(3) & 80.20(7) & 87(3) \\
       & Cl1--Ir2--Cl2 & 77.8(9) & 80(1) & 80(1) & 89(3) & 80.20(7) & 87(3) \\
       & Ir1--Cl1--Ir2 & 79.0(8) & 78(1) & 77(1) & 78(3) & 77.30(6) & 77(3)\\
       & Ir1--Cl2--Ir2 &	79.9(8) & 80(1) & 77.8(9) & 78(2) & 77.50(6) & 77(2) \\
        \hline \hline
    \end{tabular*}
\end{table*}

\captionsetup[table]{labelfont=bf,textfont=normalfont}

\begin{table*}[!ht]
    \caption{The interatomic distances and bond angles of \ce{[Rh(COD)Cl]2} at the start of the XRD experiment and at a dose of 34.8~MGy, for the 1:2, 1:6 and 1:10 light:dark X-ray radiation regimes.}\label{tab:Rhbonds}
    \centering
    \begin{tabular*}{0.79\textwidth}{l|c|ll|ll|ll}
        \hline \hline
       & \multirow{2}{*}{Bond} & \multicolumn{2}{c|}{1:2} &  \multicolumn{2}{c|}{1:6} & \multicolumn{2}{c}{1:10} \\
       & & start & end  & start & end  & start & end \\
        \hline
        \multirow{6}{*}{Length (\AA)} & Rh1--Rh2 & 3.525(3) & 3.543(3) & 3.525(3) & 3.540(3) & 3.530(2) & 3.539(3) \\
        & Cl1--Cl2 & 3.167(8) & 3.166(8) & 3.295(8) & 3.271(8) & 3.285(7) & 3.267(8)\\
        & Rh1--Cl1 & 2.398(6) & 2.414(6) & 2.431(6) & 2.449(6) & 2.42(5) & 2.422(6)\\
        & Rh1--Cl2 & 2.374(6) & 2.387(6) & 2.424(6) & 2.424(6) & 2.450(6) & 2.439(6)\\
        & Rh2--Cl1 & 2.380(6) & 2.364(6) & 2.434(6)& 2.403(6) & 2.413(5) & 2.364(6)\\
        & Rh2--Cl2 & 2.345(6) & 2.355(6) & 2.376(6) & 2.378(6) & 2.380(5) & 2.421(6)\\
        \hline
        \multirow{4}{*}{Angle ($^{\circ}$)} & Cl1--Rh1--Cl2 & 83.1(2) & 82.5(2) & 85.5(2) & 84.3(2) & 84.9(2) & 84.4(2) \\
        & Cl1--Rh2--Cl2 & 84.2(2) & 84.3(2) & 86.5(2) & 86.3(2) & 86.5(2) & 86.1(2)\\
        & Rh1--Cl1--Rh2 & 95.1(2) & 95.7(2) & 92.8(2) & 93.7(2) & 93.8(2) & 93.9(2) \\
        & Rh1--Cl2--Rh2 &	96.7(2) & 96.7(2) & 94.5(2) & 95.0(2) & 93.9(2) & 94.9(2)\\
        \hline \hline
    \end{tabular*}
\end{table*}

\begin{figure*}[h]
\centering
\includegraphics[width=\textwidth]{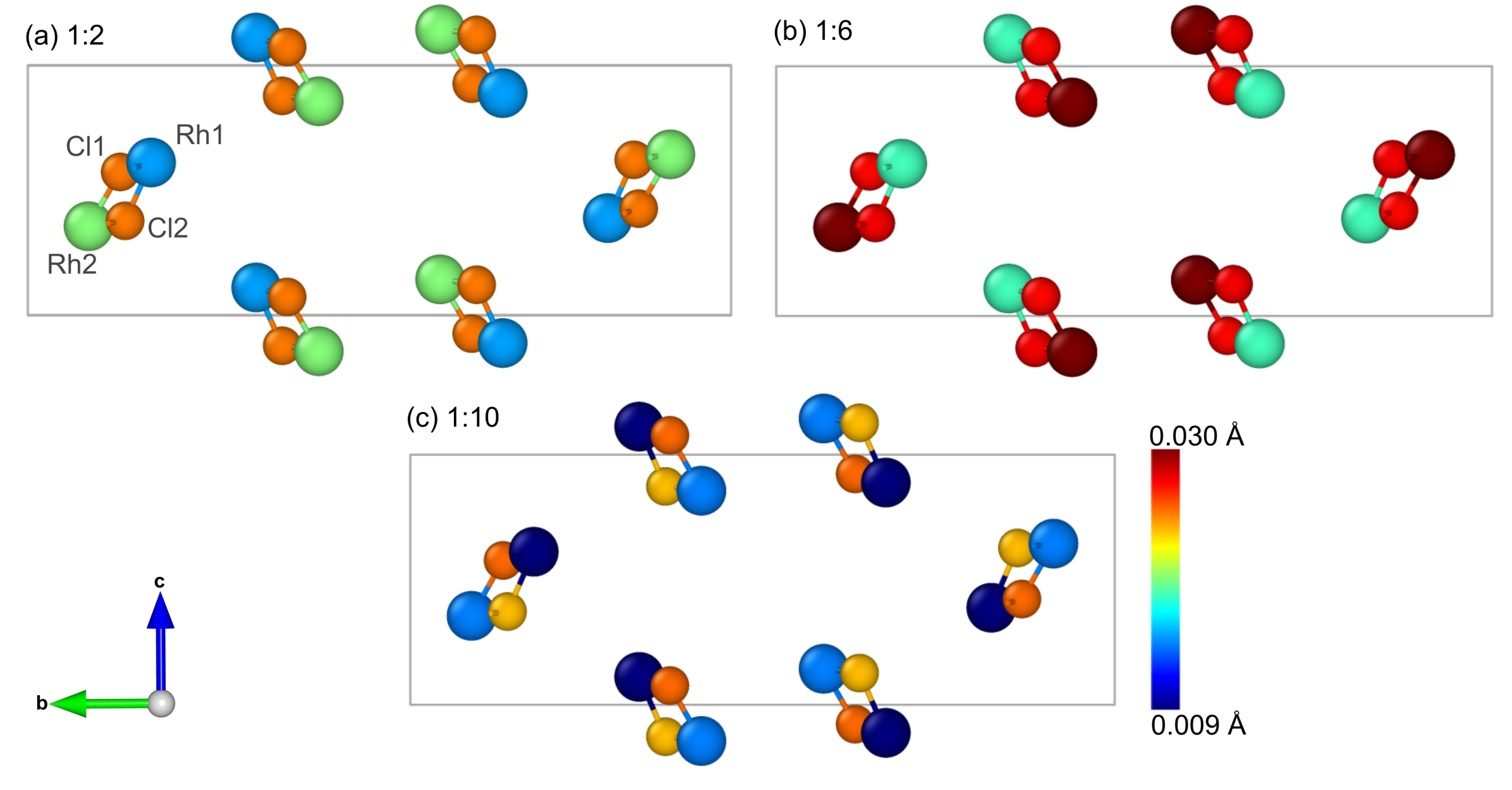}
\caption{\label{fig:Rh_rietstructure}The magnitude of atomic displacements, in \AA, of the central M--Cl core of the \ce{[Rh(COD)Cl]2} unit cell after an absorbed dose of 35~MGy, relative to the minimum dose structure for the three regimes of X-ray irradiation to X-ray-free ratios of (a) 1:2, (b) 1:6 and (c) 1:10. Figure produced using the Ovito software package.~\cite{ovito2010}} 
\end{figure*}

\begin{figure}[h]
    \centering
    \includegraphics[width=\textwidth]{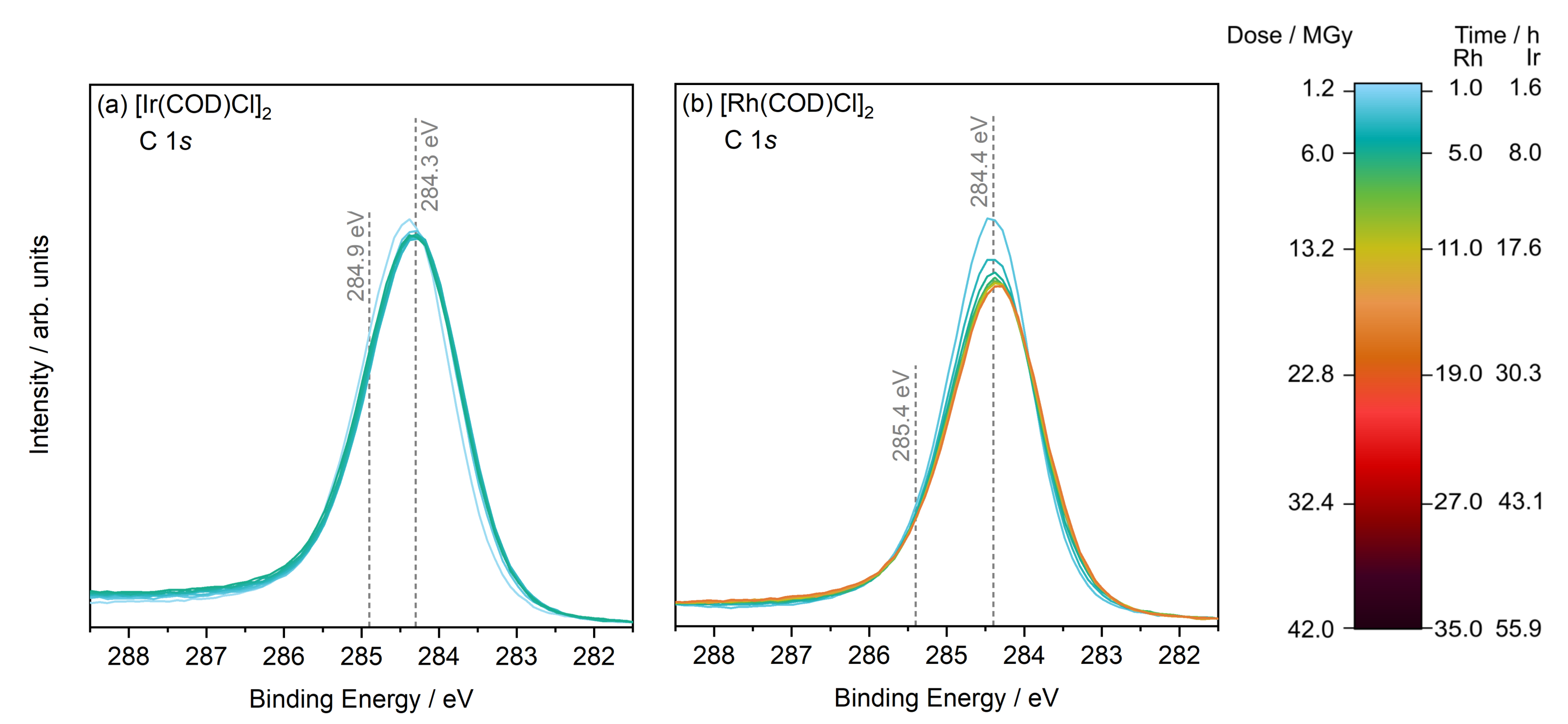}
	\caption{The iterative X-ray photoelectron C~1\textit{s} spectra of (a) \ce{[Ir(COD)Cl]2} and (b) \ce{[Rh(COD)Cl]2} collected up to a maximum absorbed dose of 11~MGy and 19~MGy, respectively, when X-ray free gaps are incorporated into the experiment.}
	\label{fig:C1s_gap_iteration}
\end{figure}

\begin{figure*}[ht]
\centering
\includegraphics[width=0.8\textwidth]{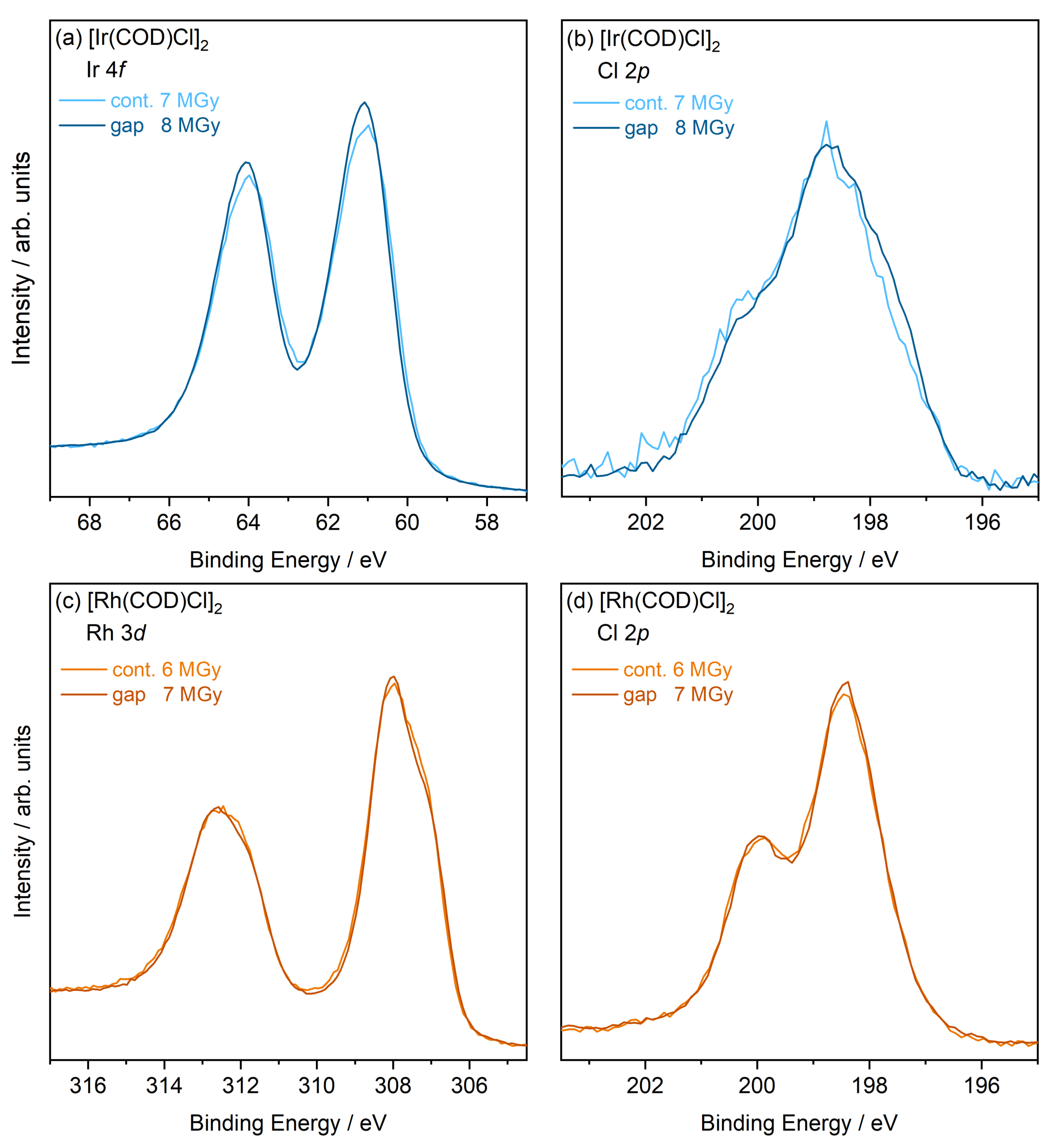}
\caption{\label{fig:XPS_cont_gap_comp}Comparison of the core level X-ray photoelectron spectra of the \ce{[M(COD)Cl]2} catalysts measured under continuous irradiation and with X-ray-free gaps at $\sim$equal dose ($\pm$1~MGy), including (a) Ir 4\textit{f} and (b) Cl 2\textit{p} spectra of \ce{[Ir(COD)Cl]2} and (c) Rh~3\textit{d} and Cl 2\textit{p} for \ce{[Rh(COD)Cl]2}. The spectra are normalised to the area of the metal.}
\end{figure*}

\clearpage

\bibliography{ref_SI} 
\bibliographystyle{abbrv}